\documentclass[
	reprint,
	superscriptaddress,
	showkeys,
	aps,
	pra,
	longbibliography,
	preprintnumbers, 
    floatfix,
	nofootinbib
]{revtex4-2}

\usepackage{lipsum}
\usepackage{subfigure}
\usepackage{mathtools}
\usepackage{amsfonts}
\usepackage{comment}
\usepackage{bbm} 
\usepackage{bm} 
\usepackage{amsmath}
\usepackage{braket}
\usepackage{textgreek}
\usepackage{xspace} 
\usepackage{graphicx}
\usepackage{color}
\usepackage{colortbl}
\usepackage{array}
\usepackage[dvipsnames]{xcolor}
\usepackage{hyperref}
\usepackage[nameinlink]{cleveref} 
\usepackage{xifthen} 
\usepackage{soul} 
\usepackage{xcolor}
\usepackage{enumitem}
\usepackage{units}
\usepackage[utf8]{inputenc} 

\newcommand{\Tr}{\text{Tr}} 
\newcommand{\tr}{\text{tr}} 
\newcommand{\vac}{\text{vac}}
\newcommand{\HFB}{\text{HFB}}
\newcommand{\NLO}{\text{NLO}}

\newcommand{\sourceJ}{J}
\newcommand{\sourceR}{R}
\newcommand{\SEcc}{\Omega}
\newcommand{\GammaInt}{\Gamma^\text{int}}
\newcommand{\nnc}{n_\text{nc}}
\newcommand{\tnnc}{\tilde n_\text{nc}}
\newcommand{\nncFree}{\nnc^0}
\newcommand{\nncInt}{n^\text{int}_\text{nc}}
\newcommand{\lambdaTh}{\lambda_\text{th}}
\newcommand{\UVco}{\Lambda_\text{UV}}
\newcommand{\IRco}{\Lambda_\text{IR}}
\newcommand{\SigmaNLO}{\Sigma_\NLO}
\newcommand{\tildeSigmaNLO}{\tilde\Sigma_\NLO}
\newcommand{\UVasymptotics}{\overset{\text{UV}}{\,\longrightarrow\,}}
\newcommand{\sWave}{$s$-wave\xspace}

\newcommand{\gettitle}{Renormalised thermodynamics for Bose gases from low to critical temperatures}
\newcommand{\getHeidelbergAffiliation}{\affiliation{Institut f\"ur Theoretische Physik, Universit\"at Heidelberg, Philosophenweg 16, 69120 Heidelberg, Germany}}
\hypersetup{ 
	colorlinks, 
	linkcolor={red!75!black}, 
	citecolor={blue!75!black},
	urlcolor={blue!75!black}, 
	pdftitle={\gettitle}, 
	pdfauthor={Heinrich},
	pdfkeywords={XXXX}
	bookmarksopen=true, 
	bookmarksopenlevel=2, 
	bookmarksnumbered=true 
}

\begin{document}
    \title{\gettitle}
    
	\author{Michael H. Heinrich}
	\email{heinrich@thphys.uni-heidelberg.de}
	\getHeidelbergAffiliation
 
	\author{Alexander Wowchik}
	\getHeidelbergAffiliation

    \author{Jürgen Berges}
	\getHeidelbergAffiliation
	
    \begin{abstract}
    We compute thermodynamic properties of dilute Bose gases using non-perturbative approximations of the two-particle irreducible (2PI) effective action. It is shown how to systematically renormalise the self-consistent descriptions beyond conventional Gaussian approximations such as Hartree-Fock-Bogoliubov theory. This allows us to determine the condensate depletion from low to high temperatures, including its critical behaviour at the phase transition. While the universal anomalous dimension at criticality is vanishing for Gaussian approximations, we determine its non-zero value at next-to-leading order of a self-consistent expansion in the number of field components.   
    \end{abstract}
    \maketitle 

\section{Introduction}
Self-consistent approximation schemes based on two-particle irreducible (2PI) effective actions provide an important means for our understanding of \mbox{(non-)equilibrium} quantum field theories~\cite{andersen2004theory, berges2021qcdThermalization}. A remarkable property is their renormalisability, since self-consistent approximations involve selective summations of infinite perturbative orders~\cite{berges2005nonperturbativeRenormalization}. While renormalisation of 2PI effective actions has been worked out and applied to the thermodynamics of relativistic field theories in detail~\cite{berges2005renormalisedThermo}, a systematic analysis of non-relativistic theories beyond Gaussian approximations like the Hartree-Fock-Bogoliubov (HFB) theory~\cite{zhang2013conserving}
is much less developed.

In this work, we show how to compute renormalised thermodynamic quantities 
for a non-relativistic scalar field theory beyond self-consistent Gaussian approximations. As an example, we consider an expansion of the 2PI effective action in the number of field components $N$ to next-to-leading order (NLO)~\cite{berges2002controlled, aarts2002FFEfrom1N, berges2005bec}. 
Applied to a dilute Bose gas, this non-perturbative approximation allows us to compute important thermodynamic quantities, such as the temperature-dependence of the condensate fraction from low to the critical temperature at the phase transition. Near criticality, fluctuations become large such that perturbative approaches fail for Bose gases~\cite{andersen2004theory}. Self-consistent approximation schemes are particularly well suited in this case, requiring no additional infrared regularisation as in alternative schemes~\cite{arnold2000tcLargeN, baym2001transitionTemperatureBose}. In particular, we present a determination of the anomalous dimension at criticality, whose non-zero value is crucial for the characterisation of the universality class of the Bose gas and which vanishes identically for Gaussian approximations.

By employing 2PI effective action techniques, both the condensate and the fluctuations around it are treated as variational parameters. Once the 2PI effective action is evaluated at the variational solution for the fluctuations, the free-energy functional (corresponding to a resummed 1PI effective action) is obtained~\cite{berges2005nonperturbativeRenormalization}. It is important that the latter is consistent with all symmetries of the Bose system, respecting conservation laws and Ward identities~\cite{reinosa2007QEDward2PI}. For instance, in Ref.~\cite{zhang2013conserving} it has been demonstrated that self-consistent Gaussian approximations based on the 2PI effective action lead in this way to a gapless excitation spectrum respecting the Hugenholtz-Pines theorem.

Gaussian approximations dramatically reduce the complexity of the resulting variational equations. In the translation-invariant case, the set of required variational parameters is reduced to two real numbers. Beyond Gaussianity the variational parameters become space-time dependent, which needs to be taken into account when the depletion of the condensate due to thermal and quantum fluctuations can no longer be neglected. The condensate and fluctuations are then obtained from variational equations for one- and two-point correlation functions, which we solve numerically as a function of temperature, density and interaction strength of the gas. 

In the presence of a Bose-Einstein condensate, additional complications arise, which do not occur in the non-condensed phase at high temperature. The variational two-point function acquires two components, one representing the single-particle propagation amplitude of non-condensed atoms, and the other the pair correlation function describing correlated scattering of particles into and out of the condensate. As a consequence, together with the condensate, three coupled self-consistent equations have to be solved instead of one equation for a single two-point correlation function. 

To obtain physical results, where the renormalised couplings are determined by their measured counterparts, these equations have to be properly renormalised. We derive the 2PI effective action renormalisation~\cite{berges2005nonperturbativeRenormalization} for the two-body coupling parameter that appears in the 
effective field theory Hamiltonian of the dilute Bose gas. We demonstrate that for the self-consistent description of the condensed phase, in addition to the standard Lippmann-Schwinger counterterm relating the physical \sWave scattering length to the coupling parameter in the ultraviolet regularised theory~\cite{braaten2006universalityFewBody, stoof2009ultracold}, a second counterterm is required at NLO beyond HFB. We explain how renormalisation is performed systematically beyond self-consistent Gaussian approximations for the Bose gas and obtain the renormalised NLO equations analytically. 

This work is organised as follows. In Sec.~\ref{sec:NLOHFBcomparison} we summarise NLO thermodynamic results for the condensate fraction, anomalous dimension and shift in the critical temperature, comparing them to those from the HFB approximation. The subsequent sections show how the results have been obtained based on the 2PI effective action for the Bose gas at non-zero temperature and density (Sec.~\ref{sec:fieldTheory}), explaining self-consistent renormalisation (Sec.~\ref{sec:renorm}) for different approximation schemes including NLO and weak coupling expansions (Sec.~\ref{sec:approx}). We end with conclusions in Sec.~\ref{sec:conc} followed by an appendix for details of the computational procedure.   

\section{Comparison of NLO and HFB results}\label{sec:NLOHFBcomparison}
\subsection{Hamiltonian and equations of motion}\label{ssec:HandEOMs}
We consider an atomic Bose gas in three spatial dimensions described by the effective field theory Hamiltonian
\begin{align} 
    \!\!\! \hat H = &\int_{\bm{x}}\! \left( -\hat\psi^\dagger (\bm{x}) \nabla_{\bm{x}}^2 \hat\psi (\bm{x})
    + \frac{g_0}{2} \hat \psi ^\dagger (\bm{x}) \hat\psi ^\dagger (\bm{x}) \hat\psi (\bm{x}) \hat\psi (\bm{x})
	\right) \!\!
\label{eq:HamiltonianWIBG}
\end{align}
with field operator $\hat\psi (\bm{x})$ and $\int_{\bm{x}}\mathord{\equiv}\int  \mathrm{d}^3 x$. Using an appropriate choice of units, we set \mbox{$\hbar=k_B=2M=1$} where $\hbar$ is the reduced Planck constant, $k_B$ is Boltzmann's constant, and $M$ denotes the atom mass. 

The ``bare'' coupling parameter $g_0$ of the Hamiltonian needs to be renormalised to relate it to the \sWave scattering length $a=g/(8\pi)$ in terms of the renormalised coupling $g$ of the theory. For the suitably regularised quantum theory, the counterterm required for the determination of the renormalised coupling has been computed and applied to approximate descriptions such as the HFB approximation to obtain physical results that are insensitive to the regulator~\cite{stoof2009ultracold, braaten2006universalityFewBody}. 

However, the HFB approximation misses important aspects of the dynamics, in particular, the non-trivial universal scaling near the critical temperature of the system. We consider the non-perturbative NLO scheme of Refs.~\cite{berges2002controlled, aarts2002FFEfrom1N, berges2005bec}, which encompasses and goes beyond HFB. We will show that thermodynamics beyond the HFB approximation generally requires an extension of the standard renormalisation procedure. In particular, at NLO a second counterterm is required to determine the renormalised behaviour in the condensed phase. 
    
As an application of this approach and to demonstrate its capabilities, we first summarise in this section the computation of the temperature dependence of the condensate fraction from low to critical temperatures. At the critical temperature, we determine the anomalous dimension, which is non-zero for the Bose gas~\cite{baym2001transitionTemperatureBose,donner2007criticalityOfTrappedBoseGas}. We compare our results with the Hartree-Fock-Bogoliubov approximation, for which the anomalous dimension vanishes identically.  
\newcommand{\downScale}{0.9}
\begin{figure*}
	\centering
    \begin{align*}
        \vcenter{\hbox{\includegraphics[scale=\downScale]{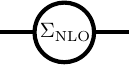}}}
        \ &= \
        \raisebox{-0.85em}{\hbox{\includegraphics[scale=\downScale]{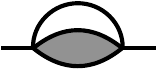}}}
        \ + \
        \raisebox{-0.85em}{\hbox{\includegraphics[scale=\downScale]{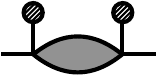}}}
        \ + \
        \raisebox{-3.09em}{\rotatebox{0}{\hbox{\includegraphics[scale=\downScale]{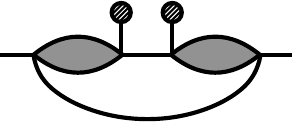}}}}\\[-2em]
        \ &= \
        \underbrace{
            \raisebox{-0.5em}{\rule{0pt}{0.1em}} 
            \raisebox{-0.2em}{\hbox{\includegraphics[scale=\downScale]{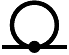}}}
            \ + \,
            \raisebox{-0.15em}{\hbox{\includegraphics[scale=\downScale]{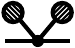}}}
        }_{
            \raisebox{1.5em}{\rule{0pt}{0.1em}} 
            \vcenter{\hbox{\includegraphics[scale=0.65]{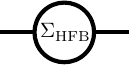}}}
        }
        \ + \
        \mathcal{O}(g^2)
    \\[-1.55em]
        \vcenter{\hbox{\includegraphics[scale=\downScale]{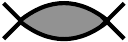}}}
        \ &= \
        \vcenter{\hbox{\includegraphics[scale=\downScale]{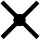}}}
        \ + \
        \vcenter{\hbox{\includegraphics[scale=\downScale]{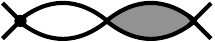}}}
        \ =
        \sum_{n=0}^{\infty}
        \
        \raisebox{-0.8em}{\hbox{\includegraphics[scale=\downScale]{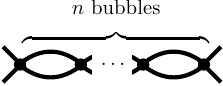}}}
        \end{align*}
        \caption{Diagrammatic contributions to the self-energy at NLO. Lines represent the propagators $G$ and $\tilde G$ while hatched circles represent insertions of the field $\Psi$, as explained in Sec.~\ref{sec:fieldTheory}. The HFB expansion is included at first order in the coupling.} 
	   \label{fig:SE_NLO_sketch} 
\end{figure*}
We consider a spatially homogeneous gas in thermal equilibrium at temperature $T$. The atomic density \mbox{$n = \braket{\hat\psi^\dagger(\bm{x}) \hat\psi(\bm{x})}$} is conserved. Below the critical temperature $T_c$, a Bose-Einstein condensate forms. In the thermodynamic limit, it is convenient to describe it in terms of the field expectation value $\Psi=\braket{\hat\psi(\bm{x})}$. We then have 
\begin{align}
	n = |\Psi|^2 + \nnc \, .
\label{eq:numCons}
\end{align}
The quantity $\Psi$ is a coherent field that captures the off-diagonal long-range order of Bose-condensation~\cite{stoof2009ultracold}, while $\nnc$ is the density of non-condensed atoms. To determine the condensate fraction $|\Psi|^2/n$ for a given macroscopic state of the gas, we find $\nnc$ and deduce $|\Psi|^2$ from~\eqref{eq:numCons}. The density of non-condensed atoms can be computed from the two-point correlation function of Heisenberg field operators $\hat\psi(x_0,\bm{x})= e^{\hat{H}x_0}\hat\psi(\bm{x})e^{-\hat{H}x_0}$ in Euclidean time $x_0$. More specifically, writing \mbox{$\hat\psi(x)\equiv \hat\psi(x_0,\bm{x})$} we consider the time-ordered two-point function
\begin{align}
	G(x-y) & = \braket{\mathcal T \hat\psi(x) \hat\psi^\dagger(y)} - |\Psi|^2 
\, , 
\label{eq:Gdef}
\end{align}
which represents the single-particle propagation amplitude of non-condensed atoms. Here $\mathcal{T}$ denotes the time-ordering operator. 
Writing $G(x-y)\equiv G(x_0-y_0;\bm{x}- \bm{y})$ one has~\cite{stoof2009ultracold}	
\begin{align}
	\nnc=\lim_{\delta x_0 \searrow 0} G\left(- \delta x_0; \bm{0}\right)
\, . 
\label{eq:nnc_fromG}
\end{align}
Additionally, we consider the pair-correlation function 
\begin{align} 
    \tilde G(x- y) &= \braket{\mathcal{T} \hat\psi(x) \hat\psi(y)} - \Psi^2 
\, ,
\label{eq:Gtdef}
\end{align}
which describes correlated scattering of particles into the condensate. Hence, $\tilde G$ vanishes for $g=0$ and for $\Psi=0$. In the following, we refer to $G$ as the \textit{normal} and to $\tilde G$ as the \textit{anomalous} propagator component.    

A central result of this work is the self-consistent computation of~\eqref{eq:Gdef} and~\eqref{eq:Gtdef} together with the condensate $\Psi$ beyond Gaussian approximations as a function of temperature and density. This is achieved by numerically solving for correlation functions in Fourier space,
\begin{align}
    G(p) = \int_x e^{ip_0 x_0 - i \bm{p} \bm{x}}\, G(x) 
\, ,
\label{eq:Gpdef}
\end{align}
where $\int_x\mathord{\equiv}\int_0^\beta \mathrm{d} x_0\int  \mathrm{d}^3 x$, and similarly for $\tilde G$. The normal propagator fulfils the equation of motion
\begin{align}
	G(p)
	&=
	\frac{ip_0 + \bm{p}^2- \mu - \Sigma^*(p)}
    {|-ip_0 + \bm{p}^2- \mu - \Sigma(p)|^2-|\tilde \Sigma(p)|^2}
\, ,
\label{eq:PE_normal}
\end{align}
where $\mu$ is the chemical potential. The respective equation of motion for the anomalous propagator reads
\begin{align}
    \tilde G(p) &=
    \frac{\tilde \Sigma(p)}
    {|-ip_0 + \bm{p}^2- \mu - \Sigma(p)|^2-|\tilde \Sigma(p)|^2}
\, .
\label{eq:PE_anom}
\end{align}
Here, the self-energies $\Sigma$ and $\tilde\Sigma$ are a consequence of interactions and depend in general on the propagators and field expectation value, 
\begin{align} 
    \Sigma=\Sigma(\Psi, G,\tilde G)
    \, , 
    \quad
    \tilde\Sigma=\tilde\Sigma(\Psi, G,\tilde G)
\label{eq:SCSE}
\, ,
\end{align}
which make~\eqref{eq:PE_normal} and~\eqref{eq:PE_anom} a coupled set of equations that is formally exact for known self-energies. 
    
In order to be able to solve these equations in practice, suitable self-energy approximations are required. Self-consistent approximation schemes that have been considered involve the HFB approximation~\cite{zhang2013conserving}. For the homogeneous system the HFB self-energies are independent of momenta and depend only on mean-field-type quantities such as $\Psi$ and $\nnc$. Within the more restrictive low-temperature Bogoliubov approximation, the self-energies $\Sigma \simeq -2g|\Psi|^2$ and $\tilde\Sigma \simeq -g\Psi^2$ are taken to depend only on the coherent field.
    
Going beyond such Gaussian approximations, we consider momentum-dependent self-energies $\SigmaNLO(p)$ and $\tildeSigmaNLO(p)$ that are obtained from the NLO scheme of Ref.~\cite{berges2002controlled, aarts2002FFEfrom1N, berges2005bec}. This non-perturbative scheme is based on an expansion in the number of field components $N$. 

We implement the expansion by considering a $U(N)$-symmetric Bose gas with $N>1$ components and classify contributions to $\Gamma[\Psi, G, \tilde G]$ according to their scaling in terms of powers of $1/N$. For $N \rightarrow 1$ the original theory given by the Hamiltonian~\eqref{eq:HamiltonianWIBG} is recovered. In this limit, $1/N$ no longer represents a small expansion parameter, and the validity of the procedure assumes continuity in the number of field components. 

The HFB approximation is included at first order in the coupling as is further described below. But for our purposes it is essential that the present non-perturbative method is not based on a coupling expansion. Due to large fluctuations, the latter would break down close to the critical temperature even for arbitrarily weak coupling. The NLO description was successfully applied before in the vicinity of second-order phase transitions~\cite{arnold2000tcLargeN, berges2004etaN}. Close to the critical point, the universal large-scale physics of the Bose gas is described by the $O(2)\sim U(1)$ universality class, which may be captured by a corresponding NLO expansion in terms of the two real components of the complex Bose field. 
    
Diagrammatically the NLO contributions are shown in Fig.~\ref{fig:SE_NLO_sketch}, which represent sums of infinitely many self-consistent contributions. While the approach is similar to widely employed ``ring'' summations, we systematically only keep the NLO contributions in the field-component expansion, which preserves the second-order nature of the phase transition. The HFB approximation formally corresponds to a truncation of $\Gamma[\Psi, G, \tilde G]$ at first order in the coupling $g_0$. The resulting self-energies are included at NLO, as depicted schematically in Fig.~\ref{fig:SE_NLO_sketch}. The contributions can be derived from the 2PI effective action $\Gamma[\Psi,G,\tilde G]$ via the variational principle
\begin{align}
    \delta\Gamma[\Psi,G,\tilde G]=0
\, ,
\label{eq:varPrinciple}
\end{align}
which implies the propagator equations~\eqref{eq:PE_normal} and~\eqref{eq:PE_anom}, along with the field equation $\delta \Gamma[\Psi,G,\tilde{G}]/\delta\Psi=0$ and~\eqref{eq:numCons}. These form a closed set of coupled equations, which are derived in sections~\ref{sec:fieldTheory} and~\ref{sec:approx}. Their solution is discussed in the following section, where we summarise results for the condensate fraction, shift in $T_c$ and the anomalous dimension. 

\subsection{Thermodynamic results from low to critical temperatures}\label{ssec:Results}
\begin{figure}
	\centering
    \includegraphics[width=0.49\textwidth]{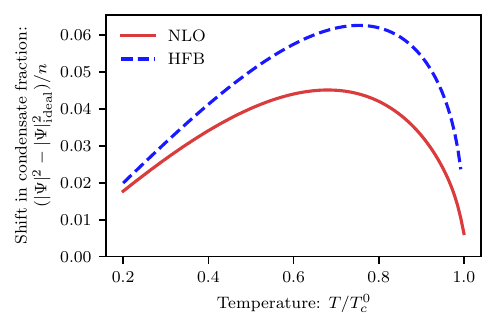}
       \caption{Shift in the condensate fraction of the interacting vs.~ideal Bose gas as a function of temperature for \mbox{$n^{1/3} a \simeq 0.004$}. The improved description of interaction effects at NLO leads to a reduced number of atoms in the condensate as compared to the HFB approximation. At low $T$ the theory becomes weakly correlated and the two approximations approach each other, while the NLO contributions are sizeable as the phase transition is approached from below.}      
\label{fig:condFrac}
\end{figure}

Our NLO result for the density of condensed atoms $|\Psi|^2$ is described as a function of temperature in Fig.~\ref{fig:condFrac} for a diluteness parameter of $n^{1/3}a =4\times 10^{-3}$. We display the difference to the ideal gas prediction, \mbox{$|\Psi|^2-|\Psi|^2_\mathrm{ideal}$}, normalised to the total density~\eqref{eq:numCons}. Here, we employ $|\Psi|^2_\text{ideal}=n-\nncFree$ with the number of non-condensed atoms $\nncFree$ in the non-interacting gas given by \mbox{$\nncFree=\zeta(3/2)\lambdaTh^{-3}$}, where \mbox{$\lambdaTh \equiv \sqrt{4\pi/T}$} is the thermal wavelength and $\zeta$ the Riemann zeta function.
The temperature in Fig.~\ref{fig:condFrac} is given in units of the critical temperature of the ideal Bose gas~\cite{andersen2004theory}
\begin{align}
    T^0_c = 4\pi \left[ \frac{n}{\zeta(3/2)} \right]^{2/3}
\, .
\label{eq:T0c}
\end{align}
This is compared to the HFB approximation, where we recover the results from Ref.~\cite{zhang2013conserving}. One observes that the HFB description over-predicts the number of condensed atoms. The difference to the NLO approximation is in principle detectable by state-of-the-art Bose gas experiments~\cite{hadzibabic2011interactionsBECtc} and expected to increase when $n^{1/3}a$ is raised further.

\begin{figure}[b]
	\centering
	\includegraphics{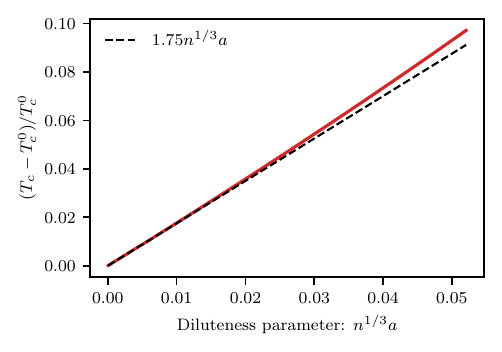}
	\caption{Shift in the critical temperature due interactions at NLO.}
\label{fig:deltaTc_plot}  
\end{figure}

In Fig.~\ref{fig:deltaTc_plot} we show how the critical temperature $T_c$ of the interacting theory at NLO changes as compared to the ideal gas prediction~\eqref{eq:T0c}. Weak interactions should reduce the critical density, such that this shift is expected to be positive with~\cite{andersen2004theory}
\begin{align}
	(T_c-T_c^0)/T_c^0 = c\, n^{1/3} a + \mathcal{O}(n^{2/3} a^2)
\, . 
\label{eq:c_def}
\end{align} 
We obtain the coefficient $c \simeq 1.75$ via the linear fit to our NLO result as displayed in Fig.~\ref{fig:deltaTc_plot}.

At the critical temperature $T_c$ of the second-order phase transition, the condensate vanishes, $|\Psi|^2=0$. Here, the correlation length diverges such that the inverse propagator vanishes at zero frequency and momentum, $G^{-1}(p = 0)=0$. The critical behaviour is described by scaling exponents. For example, the infrared scaling of the propagator at $T_c$ is given by 
\begin{align}
	\lim_{|\bm{p}| \rightarrow 0} G^{-1}(p_0=0; \bm{p}) \propto |\bm{p}|^{2-\eta}
    \, ,
    \label{eq:G_criticalScalingAnsatz}
\end{align}
where the anomalous dimension scaling exponent $\eta$ is universal. 

\begin{figure}
	\centering
	\includegraphics{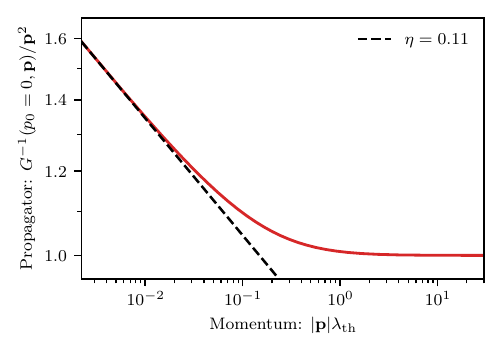}
	\caption{Critical scaling of the inverse propagator at NLO. Momenta are rescaled with the thermal wavelength $\lambdaTh$.}	
\label{fig:eta_plot} 
\end{figure}

The behaviour at NLO as a function of spatial momenta, which are rescaled with the thermal wavelength at $T=T_c$, is depicted in Fig.~\ref{fig:eta_plot}. One observes the build-up of the non-trivial scaling behaviour at low momenta. For comparison, the HFB approximation assumes a vanishing anomalous dimension and does not capture the collective phenomenon underlying the non-analytic critical behaviour of the interacting Bose gas. 

We numerically recover the result $\eta \simeq 0.11$, which was derived in Ref.~\cite{berges2004etaN} for the $O(2)$ symmetric model. Although the self-consistent approach improves the NLO estimate compared to standard $1/N$ expansion results (which use the classical propagator in loops), the very small anomalous dimension is difficult to capture, and NNLO and higher corrections are important for $N=2$~\cite{berges2004etaN}. For instance, lattice simulations of the XY model in the same universality class yield $\eta_\text{lattice} \simeq 0.038$~\cite{hasenbusch2025xy}. 

Our result for the shift in the critical temperature encoded in the coefficient $c$ of~\eqref{eq:c_def} differs by about 30\% from values obtained in lattice calculations~\cite{arnold2001BECimperfectTc, kashurnikov2001TcBEC, landau2004BECtcPIMC} and variational perturbation theory~\cite{kastening2004tcBEC7L}. The self-consistent NLO result \mbox{$c^\text{2PI}_\NLO\simeq1.75$} is remarkably close to the higher-order (NNLO) result \mbox{$c^\text{1PI}_\text{NNLO}\simeq1.71$} obtained in Ref.~\cite{arnold2000tcLargeN} based on an expansion without self-consistency. A similar scheme yields the value \mbox{$c^\text{1PI}_\NLO\simeq2.33$} at NLO~\cite{baym2000tcBECNinternalStates}. This indicates the improved expansion properties of the self-consistent approximation.    

\section{Field theory for Bose gases} \label{sec:fieldTheory}
In the following, we derive the results presented in Sec.~\ref{sec:NLOHFBcomparison} starting from field theory and consider renormalisation of self-consistent approximations in subsequent sections.  

\subsection{Path integral quantization in  equilibrium}
We employ the path integral formulation of the thermal field theory defined by the Hamiltonian~\eqref{eq:HamiltonianWIBG} and the grand canonical density operator
\begin{align}
    \hat\varrho = \frac{1}{Z} \exp \left[ - \beta(\hat H - \mu \hat N) \right]
\, ,
\label{eq:eqDm}
\end{align}
where the inverse temperature is $\beta=1/T$. For the  partition function 
\begin{align}
    Z=\Tr[  \hat\varrho] 
\, ,
\label{eq:Z}
\end{align}
we consider the path integral formulation obtained by interpreting the density matrix as a time evolution operator for imaginary times $t=[0, -i \beta)$~\cite{stoof2009ultracold} 
\begin{align}
    Z = \int \mathcal{D}\psi \, e^{-S[\psi]}
\, .
\label{eq:PathInt}
\end{align}
The classical action is given by
\begin{align}\notag
    S[\psi]=&\int_x \bigg\lbrace
    \psi^*(x) \left[\partial_{x_0} -\nabla^2_{\bm{x}}- \mu \right] \psi(x)&
\\
    &+\frac{g_0}{2} \left[\psi^*(x)\psi(x)\right]^2
    \bigg\rbrace 
\, .
\label{eq:Sclassical}
\end{align}
The path integral measure 
\begin{align}\label{eq:PathIntMeasure}
\int \mathcal{D}\psi=\int \prod_{x}\left[ \mathrm{d} \psi(x)\mathrm{d} \psi^*(x)\right]\big\vert_{\psi(0,\bm{x})=\psi(\beta,\bm{x})} \,  
\end{align}
takes thermal and quantum fluctuations into account by integrating over all configurations of the fluctuating field $\psi(x)\equiv\psi(x_0,\bm{x})$. The fluctuating field is complex-valued and periodic in imaginary time.

It is convenient to employ an index-notation where $(\psi^1, \psi^2)\mathord{\, \equiv\,}(\psi, \psi^*)$. For the fields $\psi^{\alpha}$ with $\alpha=1,2$ complex conjugation then corresponds to multiplication with the first Pauli matrix, $\left(\psi^{\alpha}\right)^*=\sigma_1^{\alpha\beta}\psi^\beta$, where summation over repeated indices is implied. With this notation for
~\eqref{eq:Sclassical}, we write
\begin{align}
    S[\psi] = S_0[\psi] + S^{\text{int}}[\psi]\, ,
\label{eq:Ssplit}
\end{align}
where the quadratic part
\begin{align}
    S_0[\psi] 
    =&\, \frac{1}{2}\int_{xy}
    \left[
    \psi^\alpha(x) \left(\mathcal{G}_0^{-1}\right)^{ \alpha \beta}(x,y) \psi^\beta(y)
    \right]
\label{eq:S0}
\end{align}
is expressed in terms of the inverse Green's function
\begin{align}
    \left(\mathcal{G}_0^{-1}\right)^{ \alpha \beta}\!(x,y)
    =
    \left[
    -i\sigma_2^{\alpha \beta}\partial_{x_0}-
    \sigma_1^{\alpha \beta}
    \left(\nabla^2_{\bm{x}}+\mu\right)
    \right] \delta(x-y) 
\, , 
\label{eq:G0inv}
\end{align}
with $\sigma_2$ the purely imaginary second Pauli matrix, and the interaction part of the classical action reads
\begin{align}\label{eq:Sint}
    S^{\text{int}}[\psi] =&\,  
    \frac{g_0}{8} \int_{x}
    \left[\psi^{\alpha}(x)\sigma_1^{\alpha\beta}\psi^\beta(x)\right]^2
    .
\end{align}
The partition function~\eqref{eq:PathInt} is promoted to a generating functional for correlation functions by introducing auxiliary source fields through
\begin{align}
    Z[J,R] \notag
    =&\,
    \int \mathcal{D}\psi \, 
    \text{exp} \bigg\lbrace
    -S[\psi] + \int_{x} \sourceJ^\alpha(x) \psi^\alpha(x) 
\\
    &+\frac{1}{2}\int_{xy} \psi^\alpha(x)\sourceR^{\alpha \beta}(x,y) \psi^\beta(y) 
    \bigg\rbrace
\, .
\label{eq:ZW_sources} 
\end{align}
For the purpose of constructing the 2PI effective action, we have introduced in addition to the linear source $\sourceJ$ also a bilinear source $\sourceR$, which satisfies $\sourceR^{\alpha\beta}(x,y)=\sourceR^{\beta\alpha}(y,x)$ by definition. By taking source-derivatives of the logarithm of the partition function, 
\begin{align}
    W[\sourceJ,\sourceR] 
    =&\,  
    \ln Z[\sourceJ, \sourceR] 
\, ,
\label{eq:Wdef}
\end{align}
we extract the connected correlation functions. In the absence of sources, one obtains the free energy density $\mathcal{F} = -W[0,0]/\beta V$~\cite{andersen2004theory}. 

The one-point function or macroscopic field is
\begin{align}
    \Psi^\alpha (x) 
    =
    \frac{\delta W[\sourceJ,\sourceR]}{\delta \sourceJ^\alpha (x)} 
    \stackrel{J=0=R}{=}
    \braket{\hat\psi^\alpha (x)} \, ,
\label{eq:PsiDef}
\end{align}
which corresponds to the field expectation value if evaluated for vanishing sources.
The number of Bose-condensed atoms is given by $|\Psi(x)|^2$, and the field expectation value can be identified with the order parameter field for the Bose-Einstein transition. 
Higher-order correlation functions are obtained from further source-derivatives. 
In particular, the description of non-condensed atoms employs the connected two-point function or propagator of the interacting quantum theory,
\begin{align}
    \mathcal{G}^{\alpha \beta }(x,y)
    &=
    \frac{\delta^2 W[\sourceJ,\sourceR]}{\delta \sourceJ^\alpha (x)\delta \sourceJ^\beta (y)}
\notag
\\
    &\!\!\!\!\stackrel{J=0=R}{=}
    \braket{\mathcal{T}\hat\psi^\alpha (x)\hat\psi^\beta (y) }
    - \Psi^\alpha (x) \Psi^\beta (y)
\, ,
\label{eq:Gcc_def}
\end{align}
which is automatically time-ordered in the path integral formulation~\cite{stoof2009ultracold}. 
As a consequence of time-ordering and the commutativity of partial derivatives, the propagator is symmetric:
\begin{align}
	\mathcal{G}^{\alpha \beta }(x,y)=\mathcal{G}^{\beta \alpha }(y,x)
    \, .
\label{eq:Gcc_symmetry}
\end{align}
Similarly, by differentiating with respect to the bilinear source $\sourceR$, one gets
\begin{align}
    \Psi^\alpha (x) \Psi^\beta (y)+\mathcal{G} ^{\alpha\beta} (x,y) 
    =
    2\frac{\delta W[\sourceJ,\sourceR]}{\delta \sourceR^{\alpha \beta} (x,y)}
\, .
\label{eq:KgeneratesG}
\end{align}

\subsection{Two-particle irreducible effective action}\label{ssec:2PI}
The 2PI effective action is the Legendre transform of $W[\sourceJ,\sourceR]$ with respect to both sources,
\begin{align} \notag
    \Gamma[\Psi, \mathcal{G}] 
    =&\, 
    -W[\sourceJ, \sourceR] 
    + 
    \int_{x}\frac{\delta W[\sourceJ,\sourceR]}{\delta \sourceJ^\alpha(x)}J^\alpha(x) 
\\
    &+ 
    \int_{x,y} \frac{\delta W[\sourceJ,\sourceR]}{\delta \sourceR^{\alpha \beta} (x,y)} 
    R ^{\alpha\beta} (x,y)
\, .
\label{eq:DoubleLegendre}
\end{align}
Consequently, the sources $J$ and $R$ on the right-hand side are chosen such that~\eqref{eq:PsiDef} and~\eqref{eq:KgeneratesG} hold. 

For vanishing sources $J=0=R$, for which physical expectation values are obtained, one finds from the 2PI effective action the variational principle:
\begin{align}
    \frac{\delta \Gamma[\Psi, \mathcal{G}]}{\delta \Psi^\alpha(x)}
    &= 0
\, , 
\label{eq:2PIvariationalPrinciple_Psi}
\\ 
    \frac{\delta \Gamma[\Psi, \mathcal{G}]}{\delta \mathcal{G}^{\alpha\beta} (x,y)}
    &= 0
\, .
\label{eq:2PIvariationalPrinciple_G}
\end{align}
To ease the notation, we will often not distinguish between the variables $\Psi$, $\mathcal{G}$ and their solution according to~\eqref{eq:2PIvariationalPrinciple_Psi},~\eqref{eq:2PIvariationalPrinciple_G}.  

We employ the interaction parametrisation of the 2PI effective action\footnote{We make use of the notation $\Tr[\mathcal{G}]\equiv\int_x \mathcal{G}^{\alpha\alpha}(x,x)$.} 
\begin{align}
    \Gamma[\Psi,\mathcal{\mathcal{G}}] 
    &=
    S_0[\Psi] 
    + 
    \frac{1}{2} \Tr [ \ln \mathcal{G}^{-1} ]
    +
    \frac{1}{2} \Tr [\mathcal{G}_0^{-1} \mathcal{G} ] + \GammaInt[\Psi, \mathcal{G}]
\, ,
\label{eq:2PIEA}
\end{align}
where $S_0$ is the non-interacting part of the classical action given by~\eqref{eq:S0}, such that $\GammaInt[\Psi,\mathcal{G}]$ contains the classical contribution~\eqref{eq:Sint} and all further coupling-dependent corrections of the quantum theory~\cite{luttingerWard1960,cornwallJackiwTomboulis1974effectiveActionForComposite}.

Plugging this parametrisation into the field equation~\eqref{eq:2PIvariationalPrinciple_Psi} yields 
\begin{align}
    \left(-i\sigma_2^{\alpha\beta}\partial_{x_0} -\sigma_1^{\alpha\beta}\left[\nabla_{\bm{x}}^2 - \mu\right] \right) 
    \Psi^\beta(x)
    + \frac{\delta \GammaInt[\Psi,\mathcal{G}]}{\delta \Psi^\alpha(x)}
    = 0 
\, ,
\label{eq:FE}
\end{align}
where the $\delta \GammaInt/\delta \Psi^\alpha$ term quantifies the impact of interactions on the condensate. By approximating $\GammaInt[\Psi,\mathcal{G}] \simeq S^\text{int}[\Psi]$, one recovers the Euclidean version of the Gross-Pitaevskii equation. If quantum corrections to $\GammaInt[\Psi,\mathcal{G}]$ are included, these are {$\mathcal{G}$-dependent} such that interactions between condensed and non-condensed atoms enter the description of the condensate. 

Moving on to the two-point function $\mathcal{G}$, its equation of motion~\eqref{eq:2PIvariationalPrinciple_G} reads 
\begin{align}
    \left(\mathcal{G}^{-1}\right)^{ \alpha \beta}(x,y) 
    &= \left(\mathcal{G}_0^{-1}\right)^{ \alpha \beta}(x,y) 
    - \SEcc^{\alpha\beta}(\Psi,\mathcal{G};x,y)
\, ,
\label{eq:PE}
\end{align}
with the self-energy
\begin{align}
    \SEcc^{\alpha\beta}(\Psi, \mathcal{G};x,y) = -2 \frac{\delta \GammaInt[\Psi,\mathcal{G}]}{\delta \mathcal{G} ^{\alpha\beta}(x,y)}
\, .
\label{eq:SEcc_def}
\end{align} 
Contributions to $\GammaInt[\Psi,\mathcal{G}]$ consist of propagators and fields contracted with the interaction vertex 
\begin{align} 
    S_{4}^{\alpha\beta\gamma\delta}(x,y,z,u) 
    =&\, 
    \frac{\delta^4 S[\psi]}{
	   \delta \psi^\alpha(x)
	   \delta \psi^\beta(y)
	   \delta \psi^\gamma(z)
	   \delta \psi^\delta(u)
	} \hspace{3em}
\notag
\\[0.35em] 
    =&\, g_0 (
    	\sigma_1^{\alpha\beta}\sigma_1^{\gamma\delta}	
    	+
    	\sigma_1^{\alpha\gamma}\sigma_1^{\beta\delta}	
    	+
    	\sigma_1^{\alpha\delta}\sigma_1^{\beta\gamma}	
    )
\notag
\\ 
    &\times\delta(x-y)\delta(x-z)\delta(z-u).
\label{eq:S4def}
\end{align}
To concisely visualise contributions to $\GammaInt[\Psi,\mathcal{G}]$, we employ the diagrammatic representation
\begin{align}
	\Psi^{\alpha}(x)
	&\equiv 
	\vcenter{\hbox{\includegraphics{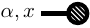}}} 
\, ,
\label{eq:Psi_diag}
\\[0.9em]
	\mathcal{G}^{\alpha\beta}(x,y)
	&\equiv 
	\vcenter{\hbox{\includegraphics{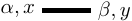}}}  
\, ,
\label{eq:G_diag}
\\[0.9em]
	S_{4}^{\alpha\beta\gamma\delta}(x,y,z,u)
	&\equiv
	\vcenter{\hbox{\includegraphics{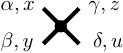}}}
\, .
\label{eq:S4_diag}
\end{align}
As an example, the contributions up to one-loop are 
\begin{align}
	\GammaInt[\Psi,\mathcal{G}]
	=&\, 
    S^\text{int}[\Psi]
    +
    \frac{g_0}{2}
	\int_{x} \bigg\lbrace
	\mathcal{G}^{\alpha\beta}(x,x)
	\Big(
	\sigma_1^{\alpha\beta} |\Psi(x)|^2
\notag
\\ 
    &+ 
	\big[\Psi^{\alpha}(x)\Psi^{\beta}(x)\big]^* 
	\Big)
	\bigg\rbrace
    + \ldots 
\notag 
\\
    =&\,
    \frac{1}{4!}
    \vcenter{\hbox{\includegraphics[scale=\downScale]{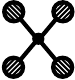}}}
    +
	\frac{1}{4}
	\raisebox{-1.65em}{\hbox{\includegraphics[scale=\downScale]{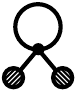}}}
    + \ldots 
\label{eq:ExampleDiagram}
\end{align}
where the dots indicate higher-loop corrections. The propagator $\mathcal{G}$ in the loops is determined self-consistently by~\eqref{eq:PE}, such that a finite number of contributions to $\GammaInt[\Psi,\mathcal{G}]$ can sum infinite perturbative loop orders in terms of the classical propagator $\mathcal{G}_0$.
In general, $\GammaInt[\Psi,\mathcal{G}]$ consists of all two-particle irreducible diagrams, which can be drawn using the components~\eqref{eq:Psi_diag}--\eqref{eq:S4_diag}. Here, two-particle irreducibility refers to the fact that a diagram cannot be disconnected by cutting two propagator lines~\cite{luttingerWard1960,cornwallJackiwTomboulis1974effectiveActionForComposite}.

Diagrammatically, computing the self-energy via the propagator derivative in~\eqref{eq:SEcc_def} amounts to cutting a single line in the diagrams in $\GammaInt[\Psi,\mathcal{G}]$. For example, the self-energy corresponding to the one-loop term in~\eqref{eq:ExampleDiagram} is given by
\begin{align} 
	&\SEcc_\text{1L}^{\alpha\beta}(x,y)
    =
	-\frac{1}{2}\ \raisebox{-0.15em}{\includegraphics[scale=\downScale]{Diagrams/SE_psi2.pdf}}
\notag
\\ &  
    \ =
	-g_0 \delta(x-y) \Big( 
		\sigma_1^{\alpha\beta} |\Psi(x)|^2 
		\! + \!
		\big[\Psi^{\alpha}(x)\Psi^{\beta}(x)\big]^*
	\Big), 
\label{eq:SEcc_1L}
\end{align}
where no propagators are attached to external legs in the diagrammatics. Approximating the self-energy in~\eqref{eq:PE} by $\SEcc_\text{1L}$  leads to a propagator $\mathcal{G}$ acquiring the form known from Bogoliubov theory~\cite{stoof2009ultracold}, as is discussed in Sec.~\ref{sec:approx}.

\subsection{Equations of motion using translation invariance}\label{ssec:EOMtransl} 
We employ spatiotemporal translation invariance of the thermal state~\eqref{eq:eqDm} to cast the field and propagator equations~\eqref{eq:FE} and~\eqref{eq:PE} into the form described in Sec.~\ref{ssec:HandEOMs}, specifically~\eqref{eq:PE_normal} and~\eqref{eq:PE_anom}. 

Since the macroscopic field is constant due to translation invariance, $\Psi^\alpha(x)\equiv\Psi^\alpha$, one obtains from~\eqref{eq:FE}
\begin{align}
	\mu\Psi = \frac{\delta \GammaInt[\Psi,\mathcal{G}]}{\delta\Psi^*(x)} 
\, ,
\label{eq:FE_hom}
\end{align} 
where we used $(\Psi^1,\Psi^2)\equiv (\Psi,\Psi^*)$. The corresponding equation holds for the complex conjugate field and is not independent. 

Translation invariance for the two-point function implies that the solution of~\eqref{eq:2PIvariationalPrinciple_G} depends only on the difference of its arguments, $\mathcal{G}^{\alpha\beta} (x,y)\equiv\mathcal{G}^{\alpha\beta} (x-y)$. 
It is thus convenient to express it in Fourier space via
\begin{align}
    \mathcal{G}^{\alpha\beta}(p)
    & =
    \int_{x} e^{i p_0 x_0- i\bm{p}\bm{x}}\,
    \mathcal{G}^{\alpha\beta}(x)
\, .
\label{eq:Gp}
\end{align}
Here $p_0$ takes on discrete values due to the periodicity of the path integral measure~\eqref{eq:PathIntMeasure}, such that $p_0 = 2\pi n T$ with $n \in \mathbb{Z}$. Furthermore, in the employed natural units the spatial momentum $\bm{p}$ corresponds directly to the wave vector. 
For momentum integrals, we employ the notation $\int_{p} =  (2 \pi) ^{-3} \int \mathrm{d}^3 \bm{p} \, T\sum_{p_0}$, where the summation $T \sum_{p_0} f(p_0)\equiv T \sum_{n=-\infty}^{n=\infty} f(2\pi T n)$ runs over all frequencies.
 
By means of complex conjugation, only two of the four components $\mathcal{G}^{\alpha\beta}$ are independent, and in momentum space one has
\begin{align}
	\big[\mathcal{G}^{21}(p)\big]^*=\mathcal{G}^{12}(p)
\, ,\quad
	\big[\mathcal{G}^{22}(p)\big]^*=\mathcal{G}^{11}(p)
\, .
\label{eq:Gp_redundancy}
\end{align}
This motivates the definitions~\eqref{eq:Gdef} and~\eqref{eq:Gtdef} of the normal and anomalous propagators 
\begin{align}
	G(x,y) \equiv \mathcal{G}^{12}(x,y)
\, ,
\quad
	\tilde G(x,y) \equiv \mathcal{G}^{11}(x,y)
\, ,
\label{eq:G_and_Gt_fromGcc}
\end{align}
which we use as the independent components of the two-point function. 
In addition to the complex conjugation relations~\eqref{eq:Gp_redundancy}, they are also subject to the symmetry~\eqref{eq:Gcc_symmetry}, which implies
\begin{align} 
    G^*(p)=G(-p), 
    \quad
    \tilde G(p)= \tilde G (-p)
\, . 	
\label{eq:G_and_Gt_symmetry}
\end{align}
Similarly, we define the normal and anomalous components of the self-energy~\eqref{eq:SEcc_def} by
\begin{align}\label{eq:SE_and_SEt_fromSEcc}
	\Sigma(x,y) \equiv \SEcc^{21}(x,y)
    \, ,
	\quad 
	\tilde\Sigma(x,y) \equiv \SEcc^{22}(x,y)
    \, .
\end{align}
Combining this with~\eqref{eq:G_and_Gt_fromGcc} and the independent component of the inverse classical propagator~\eqref{eq:G0inv}, 
\begin{align}
    G^{-1}_0 (p) = - i p_0 + \bm{p}^2 - \mu 
\, ,
\label{eq:G0p_inv}
\end{align} 
the propagator equation~\eqref{eq:PE} can be rearranged into~\eqref{eq:PE_normal} and~\eqref{eq:PE_anom}.

To proceed, it is useful to adapt the diagrammatic notation to the individual complex components which we have discussed here. For $x_0>y_0$, $G(x,y)$ represents the probability amplitude of a gas atom propagating from point $y$ to point $x$. 
In contrast, the anomalous propagator $\tilde G(x,y)$ encodes the process of a pair of atoms scattering into the Bose-Einstein condensate. Consequently, the functions $\tilde G$ and $\tilde \Sigma$ vanish in the absence of a condensate. 

We translate the diagrammatic notation~\eqref{eq:G_diag} to individual propagator components by employing
\begin{align}
    \Psi^\alpha(x)
    &\equiv 
    \left(\vcenter{\hbox{\includegraphics{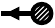}}}\, , \ \vcenter{\hbox{\includegraphics{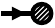}}}\right),
\label{eq:Psicc_diag}
\\[0.5em]
    \mathcal{G}^{\alpha\beta}(x,y)
    &\equiv
    \begin{pmatrix}
        \includegraphics{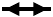} &
        \includegraphics{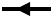} \\[0.25em]
        \includegraphics{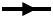} &
        \includegraphics{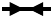}
    \end{pmatrix}.
\label{eq:Gcc_diag}
\end{align}
\newline
Furthermore, we represent the self-energy components via 
\begin{align}
	\Sigma(x,y) 
	&\equiv \vcenter{\hbox{\includegraphics[scale=0.95]{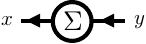}} } 
\, , 
\label{eq:SE_def} 
\\[0.3em] 
	\tilde\Sigma(x,y) 
	&\equiv \vcenter{\hbox{\includegraphics[scale=0.95]{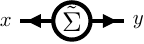}} }  
\, . 
\label{eq:SEt_def} 
\end{align}
In terms of the notation of equations~\eqref{eq:Psicc_diag} and~\eqref{eq:Gcc_diag} the vertex~\eqref{eq:S4_diag} becomes
\begin{align} 
	\vcenter{\hbox{\includegraphics{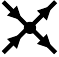}}}
    =
    2g_0 \delta(x-y)\delta(x-z)\delta(z-u)
\, .
\label{eq:S4_cc_diag}
\end{align}

\section{Vacuum renormalisation of the two-body coupling}\label{sec:renorm}
In this section, we derive the renormalisation procedure for the 2PI effective action~\cite{berges2005nonperturbativeRenormalization} for the field theory of the non-relativistic Bose gas. Renormalisation provides the connection between the parameters of the Hamiltonian, such as the bare coupling $g_0$ in~\eqref{eq:HamiltonianWIBG}, and observables like the \sWave scattering length $a$, which defines the physical two-body coupling
\begin{align}
	g = 8 \pi a 
\, .
\label{eq:gdef}
\end{align} 
Moreover, the procedure ensures that UV divergent terms drop out of calculations once the scattering length is used as input parameter to set the renormalisation condition. To this end, all expressions for quantum corrections are first regularised, where we employ a UV-cutoff $\UVco$ for momentum integrals as in
\begin{align}
	\int_{0}^{\infty} \mathrm{d} | \bm{p} | \, \rightarrow \,  \int_{0}^{\UVco} \mathrm{d} | \bm{p} | 
\, .
\label{eq:polarCoords}
\end{align}
Physical results then become insensitive to variations of $\UVco$, which we also demonstrate in appendix~\ref{app:UV}.

\subsection{Renormalisation condition in vacuum} \label{ssec:RenormConditions} 
The physical coupling~\eqref{eq:gdef} is encoded in the proper four-vertex of the quantum field theory,
\begin{align}
	\Gamma_4^{\alpha \beta \gamma \delta} (x,y,z,u) 
	= 
   \frac{\delta^4 \Gamma[\Psi, \mathcal{G}(\Psi)]}{\delta \Psi^\alpha (x) \delta \Psi^\beta (y) \delta \Psi^\gamma (z)\delta  \Psi^\delta (u) }
    \Big\vert_{\Psi=0}
\, ,
\label{eq:Gamma4def}
\end{align}
which takes into account all quantum corrections in the absence of approximations. Here $\mathcal{G}(\Psi)$ is the solution of the propagator equation~\eqref{eq:PE} and we consider $\Psi=0$. 
    
Since the physical coupling $g$ is obtained from the \sWave scattering length at low momenta, we consider the proper vertex~\eqref{eq:Gamma4def} in Fourier space, where it becomes a function of momenta. At vanishing external momenta we employ
\begin{align}
    \Gamma_4^{\alpha \beta \gamma \delta} (p=0) \equiv
    \int_{xyz} \Gamma_4^{\alpha \beta \gamma \delta}(x,y,z,0)
\, , 
\label{eq:vertexzerop}
\end{align}
where we use translation invariance to set one argument of $\Gamma_4$ in configuration space to zero. 

We evaluate~\eqref{eq:vertexzerop} in vacuum, i.e.,~at ${n=T=0}$, where we write
\begin{align}
	\Gamma_{4,\vac}^{\alpha \beta \gamma \delta} (p=0) 
    &= 
	g \left(
       \sigma_1 ^{\alpha\beta}\sigma_1 ^{\gamma\delta}
       +\sigma_1 ^{\alpha\gamma}\sigma_1 ^{\beta\delta}
       +\sigma_1 ^{\alpha\delta}\sigma_1 ^{\beta\gamma}
   \right) 
\, .
\label{eq:vacRC}
\end{align}
The tensor structure involving the first Pauli matrix is the same as for the classical action $S[\Psi]$ given by~\eqref{eq:S4def}. In particular, without any quantum corrections $\Gamma_{4,\vac}$ would equal $S_4$, and we have \mbox{$\Gamma_{4,\vac}=S_4+(\text{quantum corrections})$}. 
For given input value of the physical coupling $g$, equation~\eqref{eq:vacRC} states the renormalisation condition for our calculations.  
Conventionally, one writes the bare coupling parameter $g_0$ appearing in the classical action~\eqref{eq:S4def} in terms of the renormalised coupling $g$ and a counterterm $\delta g_0$ with
\begin{align}
    g_0 = g + \delta g_0
\, .
\label{eq:deltag0def}
\end{align}       
One then determines $\delta g_0$ as a function of $g$ and $\UVco$ using the renormalisation condition~\eqref{eq:vacRC} for any given approximation. For instance, for the vacuum of the non-relativistic Bose gas, the calculation can be done without approximations and~\eqref{eq:vacRC} leads to the well-known result~\cite{braaten2006universalityFewBody}
\begin{align}
   g_0 = \frac{g}{1- g\UVco/(4\pi^2)}
\, , 
\label{eq:g0result}
\end{align}
which can also be derived from quantum mechanics via the Lippmann-Schwinger equation~\cite{stoof2009ultracold}. 
    
The reason that this calculation can be done exactly in vacuum is, in particular, due to the lack of particle-hole processes in the absence of a medium for the non-relativistic system. 
This is very different for relativistic quantum field theories, where the vacuum state admits particle-antiparticle processes, and exact results are not available in general.
Specifically, following the discussion in section~\ref{ssec:2PI}, the exact vacuum solution to the propagator equations for the Bose gas reads 
\begin{align}
    G_\vac(p)&=\left(-ip_0+ \bm{p}^2\right)^{-1}
    ,
\label{eq:Gvac}
\end{align}
due to $\Sigma_\vac(p)=0$. At the same time, $n_\vac=0$ implies $\Psi_\vac=\tilde G_\vac(p)=\tilde \Sigma_\vac(p)=0$. The vacuum self-energy $\Sigma_\vac$ vanishes since any loop correction includes a cyclic subdiagram, which itself vanishes in vacuum:
\begin{align}
	\vcenter{\hbox{\includegraphics[scale=\downScale]{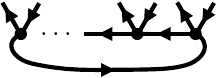}}} \big\vert_{\vac} = 0
\, .
\label{eq:cyclVacLoopsVanish}
\end{align}
In principle, renormalisation conditions can be posed in any state, not restricted to the vacuum~\cite{berges2005nonperturbativeRenormalization}. 
This includes thermal states at temperatures above or below $T_c$, where $\Psi$ can also be non-zero. 
However, some of the mentioned simplifications would then not apply. 

Although the vacuum result~\eqref{eq:g0result} is exact, thermodynamic quantities at non-zero temperature and density can typically only be determined approximately. In general, then also the renormalisation has to be done for the chosen approximation. In particular, exact results have to be adapted accordingly.    
We find that the renormalisation of self-consistent approximations can lead to a \textit{normal counterterm} $\delta g_N$ and an \textit{anomalous counterterm} $\delta g_A$, in addition to $\delta g_0$. These new counterterms arise from resummations, depending on whether the four-vertex couples to the normal propagator $G$ or the anomalous propagator $\tilde G$ in a loop diagram for quantum corrections. 
For instance, the different counterterms appear for the HFB approximation as 
\begin{align} 
    \GammaInt_\HFB[&\Psi, \mathcal{G}]
    = \int_{x} \bigg \lbrace \frac{g_0}{8}  \big[\Psi^{\alpha}(x) \sigma_1^{\alpha\beta} \Psi^\beta (x) \big]^2
\notag 
\\ 
    &+ 
    \frac{1}{4} \bigg[
        2g_N\sigma_1 ^{\alpha\beta} \sigma_1^{\gamma\delta} 
        +g_A \big(\delta^{\alpha \beta} \delta^{\gamma \delta}- \sigma_3^{\alpha \beta} \sigma_3^{\gamma \delta} \big)
    \bigg]
\notag 
\\ 
    &\times\mathcal{G} ^{\alpha\beta} (x,x)
    \left[ \Psi^\gamma(x)  \Psi^\delta(x)  
        +
        \frac{1}{2} \mathcal{G} ^{\gamma\delta} (x,x )
    \right]
    \bigg \rbrace
\label{eq:GammaHFB_cc}
\\ 
	&=\frac{1}{4}\vcenter{\hbox{\includegraphics[scale=\downScale]{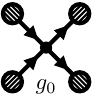}}}
    +\phantom{\frac{1}{1}}\vcenter{\hbox{\includegraphics[scale=\downScale]{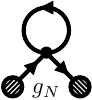}}}
    +\frac{1}{2}\raisebox{-1.8em}{\hbox{\includegraphics[scale=\downScale]{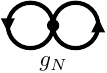}}}
\notag 
\\[0.5em]
    &+\frac{1}{4}\vcenter{\hbox{\includegraphics[scale=\downScale]{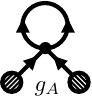}}}
    +\frac{1}{4}\vcenter{\hbox{\includegraphics[scale=\downScale]{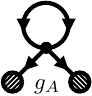}}}
    +\frac{1}{4}\,\raisebox{-1.8em}{\hbox{\includegraphics[scale=\downScale]{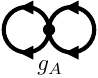}}}     
\, .
\label{eq:GammaHFB_diag}
\end{align}
Here we introduced normal and anomalous bare coupling parameters like in~\eqref{eq:deltag0def}, with
\begin{align}
    g_N &= g + \delta g_N \, , 
\label{eq:gNdef}
\\ 
    g_A &= g + \delta g_A \, . 
\label{eq:gAdef}
\end{align}
The fact that $g_N$ and $g_A$ only couple to normal and anomalous propagators, respectively, is ensured by the tensor structure in~\eqref{eq:GammaHFB_cc}. In the absence of counterterms, the tensor structure simplifies to the one found in $S_4^{\alpha \beta\gamma\delta}$ and $\Gamma_4^{\alpha \beta\gamma\delta}$ due to
\begin{align}
    &\sigma_1^{\alpha \gamma}\sigma_1^{\beta \delta}
	+\sigma_1^{\alpha \delta} \sigma_1^{\beta \gamma}
    =\sigma_1 ^{\alpha\beta} \sigma_1^{\gamma\delta} 
    +\delta^{\alpha \beta} \delta^{\gamma \delta}
    -\sigma_3^{\alpha \beta} \sigma_3^{\gamma \delta} 
    \, .
    \label{eq:GammaIntHFB_simplifiedTensorstructure}
\end{align} 
    
One then determines $\delta g_0$, $\delta g_N$, and $\delta g_A$ as a function of $g$ and $\UVco$, such that after imposing renormalisation conditions UV divergent terms drop out of physical results. 
Both $\delta g_A$ and $\delta g_N$ can be computed exactly in vacuum, where they equal $\delta g_0$, leading to~\eqref{eq:g0result}. 
However, in the presence of approximations, they differ in general. 
For the HFB approximation we will find
\begin{align}
    \delta g_{A}^{\HFB} = \delta g_0 
\, ,
\label{eq:exactBareCouplings}
\end{align}    
while the normal counterterm turns out to vanish,  
\begin{align}
    \delta g_N^\HFB=0 
\, . 
\label{eq:gNvanishesAtHFB}
\end{align}    
We will see that the result for $\delta g_N$ varies for the different approximations of $\GammaInt[\Psi, \mathcal{G}]$ we consider, while $\delta g_A$ only receives contributions at the HFB level and remains unchanged by going beyond that approximation.
 
\subsection{Vertex functions}\label{ssec:2PIvertices} 
In this section, we introduce the tools which allow us to evaluate the renormalisation condition~\eqref{eq:vacRC} using approximations of the 2PI effective action. We consider the system in the non-condensed phase where $\Psi=0$. In subsequent sections, we will then specify to the vacuum state. 

For the renormalisation condition we employ the proper vertex $\Gamma_4$, which is given by~\eqref{eq:Gamma4def} as the fourth field-derivative of the effective action $\Gamma[\Psi, \mathcal{G}(\Psi)]$. Here $\mathcal{G}(\Psi)$ is the solution of the propagator equation~\eqref{eq:2PIvariationalPrinciple_G} for a given field $\Psi$. Applying the chain rule, one finds that
\begin{align}\notag
	\Gamma_4^{\alpha \beta \gamma \delta} (x,y,z,u) \notag
	&= 
	\frac{\delta^4 \GammaInt[\Psi, \mathcal{G}]}{\delta\Psi^\alpha(x) \delta\Psi^\beta(y) \delta\Psi^\gamma(z) \delta \Psi^\delta(u)}
	\big\vert_{\bar{\mathcal{G}},\Psi=0}
\\ \notag
	&+\frac{1}{2} \int_{vw} \bigg\lbrace
	\Lambda^{\alpha\beta,\eta \kappa}(x,y;v,w)
\\
    &\times\frac{\delta^2\mathcal{G}^{\eta\kappa}(\Psi;v,w)}{\delta\Psi^\gamma(z)\delta\Psi^\delta(u)}
	\big\vert_{\Psi=0}
	+\text{perm.}
	\bigg\rbrace 
    \, , 
\label{eq:Gamma4_1}
\end{align}
where ``\text{perm.}" refers to the distinct permutations of the pairs of labels $(\alpha,x),(\beta,y),(\gamma,z),(\delta,u)$, as illustrated in~\eqref{eq:Gamma4_2diag}. 

To ease the notation, we write $\bar{\mathcal{G}}\equiv\mathcal{G}(\Psi=0)$ and
\begin{align}\label{eq:LambdaDef}
	\Lambda ^{\alpha\beta, \gamma\delta}(x,y;z,u) \equiv 2 \frac{\delta^3 \GammaInt[\Psi, \mathcal{G}]}{\delta\Psi^{\alpha}(x)\delta\Psi^{\beta}(y) \delta\mathcal{G}^{\gamma\delta}(z,u)}
	\big\vert_{\bar{\mathcal{G}}, \, \Psi=0}
\, .
\end{align}
The functions $\delta^4\GammaInt/\delta\Psi^4$ and $\Lambda^{\alpha\beta,\gamma\delta}$ are the required building blocks for the renormalisation condition~\eqref{eq:vacRC}. They can be obtained from any given approximation of the 2PI effective action~\eqref{eq:2PIEA} by taking functional derivatives of its interacting part $\GammaInt[\Psi,\mathcal{G}]$.

To proceed, we evaluate the second-order field-derivative of the propagator, which appears in the expression~\eqref{eq:Gamma4_1} of the vertex. By employing the propagator equation~\eqref{eq:PE}, we find
\begin{align} \notag
    &\frac{\delta^2 \mathcal{G} ^{\alpha\beta} (\Psi;x,y)}{\delta \Psi^\gamma (z)\delta \Psi^\delta (u)}
    \big\vert_{\Psi=0}
    \\ 
    & \hspace{1.5em}=
    -
    \int_{vw} \bar{\mathcal{G}}^{\alpha\eta}(x,v)\bar{\mathcal{G}}^{\beta\kappa}(y,w)
    V^{\alpha\beta, \gamma\delta}(v,w;z,u),
\label{eq:d2GdPs2}
\end{align}
which is given in terms of the ``2PI-vertex"
\begin{align}\label{eq:Vdef}
V^{\alpha\beta, \gamma\delta}(x,y;z,u)
\equiv
-\frac{\delta^2 \SEcc ^{\alpha\beta} \big(\Psi, \mathcal{G};x,y \big)}{\delta \Psi^\gamma (z)\delta \Psi^\delta (u)} \big\vert_{\bar{\mathcal G},\Psi=0}.
\end{align}
This four-point function $V$ is determined self-consistently by
\begin{align} \notag
	&V^{\alpha\beta, \gamma\delta}(x,y; z,u) = \,
	\Lambda^{\alpha\beta,\gamma\delta}(x,y;z,u)
    \\ \notag
    & \hspace{1em} -  \frac{1}{2}\int_{v_1v_2w_1w_2} \bigg\lbrace\Lambda^{\alpha\beta, \eta_1\kappa_1} (x,y; v_1,w_1)
	\bar{\mathcal{G}}^{\eta_1\eta_2}(v_1,v_2) 
	\\ & \hspace{1em}\times\bar{\mathcal{G}}^{\kappa_1\kappa_2}(w_1,w_2) 
	V ^{\eta_2 \kappa_2, \gamma\delta}(v_2,w_2;z,u)
	\bigg\rbrace .
    \label{eq:BSE}
\end{align}
The latter follows from the chain rule and the propagator equation~\eqref{eq:PE}, which relates $\bar{\mathcal{G}}$ with the self-energy $\SEcc$.
The expression resums terms of the form $\Lambda \bar{\mathcal{G}}^2$ and ensures that the renormalisation procedure does not miss any sub-divergences generated by the resummation implied by~\eqref{eq:PE}~\cite{berges2005nonperturbativeRenormalization}.
Diagrammatically, it reads
\begin{align}\label{eq:BSEdiag}	    
    \!\!\!\vcenter{\hbox{\includegraphics{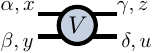}}} 
    =
    \, \vcenter{\hbox{\includegraphics{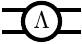}}}
    \, -\frac{1}{2}\,
    \vcenter{\hbox{\includegraphics{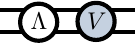}}} 
    \, .
\end{align}
To arrive at~\eqref{eq:BSE}, one also uses the relation
\begin{align}
	\frac{\delta \GammaInt[\Psi, \mathcal{G}]}{\delta \Psi^\alpha (x) \delta \Psi^\beta (y)} \big\vert_{\Psi=0} 
	= 
	2 \frac{\delta \GammaInt[\Psi, \mathcal{G}]}{\delta \mathcal{G} ^{\alpha\beta} (x,y)}
	\big\vert_{\Psi=0}
\label{eq:2kernelSimp}
\, ,
\end{align}
which implies 
\begin{align}
	\Lambda^{\alpha\beta,\gamma\delta}(x,y;z,u) =4 \frac{\delta^2 \GammaInt[\Psi, \mathcal{G}]}{ \delta\mathcal{G}^{\alpha\beta}(x,y) \delta\mathcal{G}^{\gamma\delta}(z,u)}\big\vert_{\bar{\mathcal{G}},\Psi=0}
\, .
\label{eq:4kernelSimp}
\end{align}
We stress that the relation~\eqref{eq:2kernelSimp} holds at arbitrary orders of the coupling expansion and the $1/N$ expansion, which are the approximations we consider here. 
However, it does not hold for loop truncations, which are therefore more complicated to renormalise. 
Finally,~\eqref{eq:Gamma4_1} can be rearranged into the form
\begin{align}
	&\Gamma_4^{\alpha \beta \gamma \delta} (x,y,z,u) \notag
	= 
	\frac{\delta^4 \GammaInt[\Psi, \mathcal{G}]}{\delta\Psi^\alpha(x) \delta\Psi^\beta(y) \delta\Psi^\gamma(z) \delta \Psi^\delta(u)}
	\big\vert_{\bar{\mathcal{G}},\Psi=0}
\\
	&\quad+\bigg\lbrace 
	V^{\alpha\beta,\gamma \delta}(x,y;z,w)-\Lambda^{\alpha\beta,\gamma \delta}(x,y;z,w)
	+\text{perm.}
	\bigg\rbrace
\, ,
\label{eq:Gamma4_2}
\end{align}
where the required permutations of $\Lambda$ and $V$ are illustrated diagrammatically as
\begin{align}
    \includegraphics{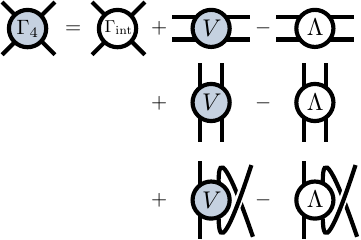} 
\ .
\label{eq:Gamma4_2diag}
\end{align}
Other permutations of external legs would not lead to distinct terms due to the symmetry property 
\begin{align}
	\Lambda^{\alpha\beta, \gamma\delta}(x,y;z,u)=
	\Lambda^{\beta\alpha, \gamma\delta}(y,x;z,u)=
	\Lambda^{\gamma\delta, \alpha\beta}(z,u;x,y)
\, ,
\label{eq:LambdaSymm}
\end{align}
which follows from~\eqref{eq:4kernelSimp} and the symmetry~\eqref{eq:Gcc_symmetry} of the propagator. By means of the self-consistent vertex equation~\eqref{eq:BSE}, it can be checked that the 2PI-vertex $V$ shares the same permutation symmetry.

In summary, after computing $\Lambda$ and ${\delta^4 \GammaInt/\delta\Psi^4}$ from the 2PI effective action, we can evaluate the renormalisation condition~\eqref{eq:vacRC} via the vertex equations~\eqref{eq:BSE} and~\eqref{eq:Gamma4_2} for the vertex $\Gamma_4$ and the 2PI-vertex $V$. 

These equations have a priori $2^4$ components. It is therefore useful to employ $U(1)$-symmetry and commutativity of partial derivatives to find the independent components of $V$ and $\Gamma_4$. Both functions are defined in the non-condensed phase, i.e. for vanishing field, $\Psi=0$. Their components, which vary under a $U(1)$ transformation, must therefore vanish, i.e., only entries of $V$ and $\Gamma_4$ with the same number of ``1" and ``2" indices are allowed. Thus, the only independent component of the proper vertex is
\begin{align} 
    \Gamma_4^{1122}(x,y,z,u)&=\vcenter{\hbox{\includegraphics[scale=0.975]{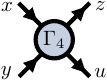}}}
\, ,
\label{eq:Gamma4_1122} 
\end{align}
and likewise for ${\delta^4 \GammaInt/\delta\Psi^4}$. For the 2PI vertex, one finds two independent components:
\begin{alignat}{2} 
    V^{12,21}(x,y;z,u) \equiv V_N(x,y;z,u)
    &=\vcenter{\hbox{\includegraphics{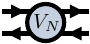}}}
\, ,
\label{eq:V_N}
\\[0.75em]  
    V^{11,22}(x,y;z,u) \equiv V_A(x,y;z,u)
    &=\vcenter{\hbox{\includegraphics{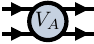}}}
\, ,
\label{eq:V_A}
\end{alignat}
and similarly for $\Lambda$. We refer to $V_N$ and $V_A$ as the normal and anomalous vertex components, since they are related to the normal and anomalous propagator derivatives through~\eqref{eq:4kernelSimp}. 

They can be computed from the corresponding components of the self-consistent vertex equation~\eqref{eq:BSE}. In the absence of anomalous propagator components, the equations for $V_N$ and $V_A$ become independent of each other and read
\begin{align} \notag
	&V_N(x,y; z,u) = 
	\Lambda_N(x,y;z,u)
\\ \notag
    &\hspace{3em} -  \int_{v_1v_2w_1w_2} \bigg\lbrace\Lambda_N (x,y; v_1,w_1)
\\ 
    &\hspace{3em} \times \bar{\mathcal{G}}^{21}(v_1,v_2) \bar{\mathcal{G}}^{12}(w_1,w_2)  V_N(v_2,w_2;z,u)
	\bigg\rbrace,
\label{eq:VN_BSE}
\\[0.5em] \notag
	&V_A (x,y; z,u) = 
	\Lambda_A(x,y;z,u)
\\ \notag
	 &\hspace{3em} -  \frac{1}{2}\int_{v_1v_2w_1w_2} \bigg\lbrace\Lambda_A (x,y; v_1,w_1)
\\& 
    \hspace{3em} \times \bar{\mathcal{G}}^{21}(v_1,v_2) 
	\bar{\mathcal{G}}^{21}(w_1,w_2) 
	V_A(v_2,w_2;z,u)
	\bigg\rbrace,
    \label{eq:VA_BSE}
\end{align}
with the diagrammatic forms
\begin{align}\label{eq:VN_BSE_diag}
	\vcenter{\hbox{\includegraphics{Diagrams/VN.pdf}}}\,
    &=
    \, \vcenter{\hbox{\includegraphics{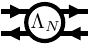}}}\,
    -
    \phantom{\frac{1}{2}}
    \ \vcenter{\hbox{\includegraphics{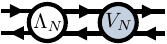}}}\,
    \raisebox{0em}{,}
	\\[1em]
	\vcenter{\hbox{\includegraphics{Diagrams/VA.pdf}}}\,
    &=
    \, \vcenter{\hbox{\includegraphics{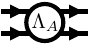}}}\,
    -\frac{1}{2}
    \ \vcenter{\hbox{\includegraphics{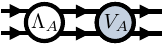}}}
    \, .
\label{eq:VA_BSE_diag}
\end{align}
The independent component of~\eqref{eq:Gamma4_2} is given by 
\begin{align}
    &\!\Gamma_4^{1122}(x,y,z,u)
    = 
    \frac{\delta^4\GammaInt[\Psi,\mathcal{G}]}{ \delta \Psi(x) \delta \Psi(y) \delta \Psi^*(z) \delta \Psi^*(u)}\big\vert_{\bar{\mathcal G},\Psi=0}
    \notag \\[0.5em]
    &\ \ + V_A(x,y;z,u) - \Lambda_A(x,y;z,u) + V_N(x,z;y,u) 
    \notag \\[0.4em]
    &\ \  - \Lambda_N(x,z;y,u) + V_N(x,u;y,z) - \Lambda_N(x,u;y,z)
\, .  
\label{eq:equation_for_Gamma4_1122}
\end{align}

Within a given approximation of the 2PI effective action, the vertex components~\eqref{eq:Gamma4_1122}--~\eqref{eq:V_A} are generally different from each other. Consequently, they approximate the full vertex~\eqref{eq:Gamma4_1122} in a different way. Indeed, it can be shown that they become equal in the exact theory~\cite{berges2005nonperturbativeRenormalization}, where $\Gamma_4^{1122}(x,y,z,u)=V_A(x,y;z,u) = V_N(x,z;y,u)$. In the presence of approximations, the renormalisation condition~\eqref{eq:vacRC} then translates into three independent conditions 
\begin{align}
    \Gamma_{4,\vac}^{1122}(p=0)=V^\vac_N(p=0) = V_A^\vac(p=0)=2g,
\label{eq:RCforIndComps}
\end{align}
which ensure proper convergence to the exact theory. 

Simultaneously fulfilling the conditions~\eqref{eq:RCforIndComps} generally demands three independent counterterms. 
In addition to $\delta g_0$, we need the normal counterterm $\delta g_N$ and the anomalous counterterm $\delta g_A$. 
In view of equations~\eqref{eq:BSE}--~\eqref{eq:Gamma4_2}, the counterterms correspond to shifts in the functions $\Lambda_N$, $\Lambda_A$ and ${\delta^4 \GammaInt/\delta\Psi^4}$. 
They can be introduced to the effective action at the level of single-vertex contributions, as shown in~\eqref{eq:GammaHFB_cc}. 

\subsection{Vacuum counterterms}\label{ssec:bareCoupl} 
We now compute the counterterms $\delta g_0$, $\delta g_N$, and $\delta g_A$ in terms of the vacuum \sWave scattering length \mbox{$a=g/(8\pi)$}.
To this end, we specify to the vacuum state given by \mbox{$T=n=\Psi=\tilde{G}=\tilde \Sigma=0$} and \mbox{$G_\vac(p)=(-i p_0 + \bm{p}^2)^{-1}$} and evaluate the renormalisation conditions~\eqref{eq:RCforIndComps} for $\Gamma_4^{1122}$,$V_N$ and $V_A$. 
Their required building blocks, $\delta^4\GammaInt/\delta \Psi^4$, $\Lambda_N$, and $\Lambda_A$ simplify considerably in vacuum as shown in appendix~\ref{app:vacuum}. 
For example, $\Lambda_A^\vac$ receives only the HFB-contribution
\begin{align}
    \Lambda_A^\vac(x,y;z,u)
    &=
    2g_A\delta(x-y)\delta(x-z)\delta(z-u) 
\, ,
\label{eq:LambdaAvac}
\\[0.9em]
    \vcenter{\hbox{\includegraphics[scale=1]{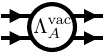}}} \
    &= \
    \raisebox{-1.5em}{\hbox{\includegraphics[scale=\downScale]{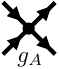}}} 
\, ,
\label{eq:LambdaAvac_diag}
\end{align} 
since all contributions with two or more vertices contain cyclic integrals that vanish in vacuum due to~\eqref{eq:cyclVacLoopsVanish}. The vertex equation~\eqref{eq:VA_BSE} then becomes
\begin{align}
    \vcenter{\hbox{\includegraphics[scale=1]{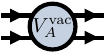}}} 
    \ = \,
    \raisebox{-1.5em}{\hbox{\includegraphics[scale=\downScale]{Diagrams/vertex_gA.pdf}}} 
    \, - \frac{1}{2}
    \ \raisebox{-1.3em}{\hbox{\includegraphics[scale=\downScale]{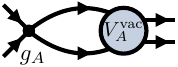}}} 
\, ,
\label{eq:VA_BSE_vac}
\end{align} 
which implies, together with the renormalisation condition~\eqref{eq:RCforIndComps}:
\begin{align}
    g=g_A-g_A \bar{P}_\vac(0) g
    .
    \label{eq:VA_BSE_result}
\end{align}
Here, we have employed the anomalous bubble integral
\begin{align}
	\bar P_\vac (p)=\int_q G_\vac(p-q) G_\vac(q) \equiv \vcenter{\hbox{\includegraphics[scale=\downScale]{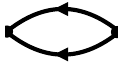}}}
\, ,
\label{eq:barPvacp}
\end{align}
whose zero-momentum part is 
\begin{align}
    \bar P_\vac (0) =  
    \frac{\UVco}{4 \pi^2}
\, .
\label{eq:todo}
\end{align}
Rearranging~\eqref{eq:VA_BSE_result}, we arrive at 
\begin{align}
	g_A = \frac{g}{1- g\UVco/(4\pi^2)}
\, , 
\label{eq:gAresult}
\end{align}
which equals the result~\eqref{eq:g0result} for $g_0$ which we find later in this section. Expression~\eqref{eq:gAresult} resums the infinite series of diagrams 
\begin{align}
    \vcenter{\hbox{\includegraphics[scale=\downScale]{Diagrams/VA_vac.pdf}}} 
    =
    \raisebox{-1.5em}{\hbox{\includegraphics[scale=\downScale]{Diagrams/vertex_gA.pdf}}}
    &-\frac{1}{2}\,\raisebox{-1.5em}{\hbox{\includegraphics[scale=\downScale]{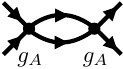}}}
    \notag\\
    &+\frac{1}{4}\,\raisebox{-1.4em}{\hbox{\includegraphics[scale=\downScale]{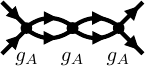}}}
    +\dots
\label{eq:VA_BSE_Bubbresum}
\end{align}
which can be obtained by iterating~\eqref{eq:VA_BSE_vac}. 

For the normal channel, the roles of $V$ and $\Lambda$ are reversed: Loop contributions appear in $\Lambda_N^\vac$, while the self-consistent equation~\eqref{eq:VN_BSE} for $V_N^\vac$ becomes trivial: 
\begin{align}
    V_N^\vac(x,y;z,u)
    &=
    \Lambda_N^\vac(x,y;z,u)
\, ,
\label{eq:VNvac}
\\[0.9em]
    \vcenter{\hbox{\includegraphics[scale=\downScale]{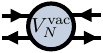}}} \
    &= \
    \vcenter{\hbox{\includegraphics[scale=\downScale]{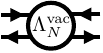}}} 
\, .
\label{eq:VNvac_diag}
\end{align}
This can be seen by iterating~\eqref{eq:VN_BSE} once,
\begin{align} \notag
    \vcenter{\hbox{\includegraphics[scale=\downScale]{Diagrams/VN.pdf}}}
    \ = \ & 
    \vcenter{\hbox{\includegraphics[scale=\downScale]{Diagrams/LambdaN.pdf}}}
    \ - \
    \vcenter{\hbox{\includegraphics[scale=\downScale]{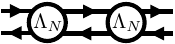}}}
\\[0.3em]
    &+ \
    \vcenter{\hbox{\includegraphics[scale=\downScale]{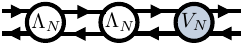}}}
\, ,
\label{eq:VN_BSE_iter}
\end{align}
which directly leads to cyclic contributions, which vanish in vacuum. Due to~\eqref{eq:VNvac} the renormalisation condition for $V_N^\vac$ applies directly to a derivative of $\GammaInt[\Psi,\mathcal{G}]$\footnote{Therefore, $\delta g_N$ plays the role of a BPHZ-counterterm, see also~\cite{berges2005nonperturbativeRenormalization}.}. 
Hence, contributions to $\GammaInt[\Psi,\mathcal{G}]$ beyond the HFB-terms~\eqref{eq:GammaHFB_cc} have to include $g_N$ as a bare coupling~\cite{berges2005nonperturbativeRenormalization}. Hence, the contributions to $\Lambda^\vac_N$ are found to be
\begin{align}
    \vcenter{\hbox{\includegraphics[scale=1]{Diagrams/LambdaN_vac.pdf}}} \, 
    &= 
    \raisebox{-1.5em}{\hbox{\includegraphics[scale=\downScale]{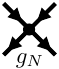}}} 
    -\frac{1}{2}\vcenter{\hbox{\includegraphics[scale=0.8]{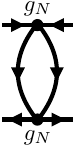}}} \, 
    +\frac{1}{4}\vcenter{\hbox{\includegraphics[scale=0.825]{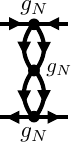}}} \, 
    + \dots
\label{eq:LambdaN_vac_exact_diagrams}
\end{align}
such that~\eqref{eq:VNvac} implies along with the renormalisation condition for $V_N$:
\begin{align}
    g &= g_N \sum_{n=0}^\infty  \big[-g_N \bar{P}_\vac(0)\big]^n 
\, .
\label{eq:gN_exact}
\end{align}
In the exact theory, all terms in the sum contribute and~\eqref{eq:gN_exact} can be brought into the form of~\eqref{eq:VA_BSE_result}, which leads to $g_N^\text{exact}=g_0$. We stress that the summation in $\Lambda^\vac_N$ does not originate from a self-consistent vertex equation, but rather from loop contributions to $\GammaInt[\Psi,\mathcal{G}]$. Therefore, the sum in~\eqref{eq:gN_exact} is truncated differently by each approximation of $\GammaInt$: One finds that $g_N$ vanishes at HFB-level and is re-adjusted beyond HFB, as further discussed in section~\ref{sec:approx}. On the other hand, $g_A$ remains the same for each approximation beyond HFB. 

Similarly to $\Lambda_A$, we find only a single-vertex contribution in
\begin{align}
	\frac{\delta^4\GammaInt[\Psi,\mathcal{G}]}{\delta \Psi(x)\delta\Psi(y)\delta\Psi^*(z)\delta\Psi^*(u)}\big\vert_{\vac}\hspace{-7.5em}&
\notag
\\[0.4em] 
    &= 2g_0\delta(x-y)\delta(x-z)\delta(z-u)
\, , 
\label{eq:GammaInt1122_vac}
\\[0.4em]
    \vcenter{\hbox{\includegraphics[scale=\downScale]{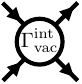}}} \
    &= \
    \raisebox{-1.5em}{\hbox{\includegraphics[scale=\downScale]{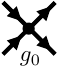}}} 
\, .
\label{eq:GammaInt1122_vac_diag}
\end{align} 
This term originates from $S_\text{int}[\Psi]$ and is therefore present at zero-loop order and unchanged by loop corrections.
Finally, we evaluate the remaining renormalisation condition~\eqref{eq:RC_for_Gamma4_1122} for $\Gamma_4$ to determine $g_0$:
Plugging~\eqref{eq:GammaInt1122_vac} and \mbox{$V_N^\vac=\Lambda_N^\vac$} into~\eqref{eq:Gamma4_1122} we arrive at
\begin{align} 
    \Gamma_{4,\vac}^{1122}(p=0) 
    =\, 
    &2g_0 + V_A^\vac(p=0)-\Lambda_A^\vac(p=0)
\, .
\label{eq:RC_for_Gamma4_1122}
\end{align} 
Using $\Lambda_A^\vac(p=0)=2g_A$ then leads to $g_0=g_A$, such that we find~\eqref{eq:exactBareCouplings}.

\section{Approximation schemes}\label{sec:approx}
In the following, we discuss the approximations of the 2PI effective action used to obtain the results of section~\ref{sec:NLOHFBcomparison}. 
We give the self-energy, the field equation and the result for the normal counterterm $\delta g_N$ at NLO in the $1/N$ expansion~\cite{berges2002controlled, aarts2002FFEfrom1N, berges2005bec}. From these we also recover the HFB approximation by dropping momentum dependent contributions~\cite{zhang2013conserving}. 

\subsection{2PI effective action expanded to NLO}
Our starting point is the $U(N)$-generalization of the $U(1)$-Hamiltonian~\eqref{eq:HamiltonianWIBG}, which reads
\begin{align}
    \hat H_{U(N)} 
    =& 
    \int \mathrm{d}^3x  \, \Big[
        -\hat\psi_a^\dagger (\bm{x}) \nabla_{\bm{x}}^2 \hat\psi_a (\bm{x})
\notag
\\
    &+ \frac{g_0}{2N} \hat \psi_a ^\dagger (\bm{x}) \hat\psi_b ^\dagger (\bm{x}) \hat\psi_a (\bm{x}) \hat\psi_b (\bm{x})
    \Big]
\, .
\label{eq:HamiltonianUN}
\end{align}
Here, the indices $a,b =1,\dots,N$ label internal (spin) degrees of freedom of the atoms.
For example, setting $N=3$ yields the Hamiltonian of a spin-1 Bose gas in the absence of spin coupling or Zeeman parameters~\cite{ueda2012spinorBECs}.
Field contractions such as $\hat\psi^\dagger_a \hat\psi_a$ are taken to scale linearly in $N$. 
Hence, we have rescaled the coupling with a factor of $1/N$, such that $\hat H_{U(N)}\sim N$. This ensures that the energy scale associated with interactions stays fixed when $N$ is varied~\cite{mikheev2019leftNTFPmultiBEC}. 
One classifies the contributions to $\Gamma[\Psi,\mathcal{G}]$ according to 
\begin{align} 
    \Gamma [\Psi,\mathcal{G}]
    &=
    \Gamma_\text{LO}[\Psi,\mathcal{G}] 
    +
    \Gamma_\NLO[\Psi,\mathcal{G}]
    + 
    \Gamma_\text{NNLO}[\Psi,\mathcal{G}]
    + \dots
\label{eq:GammaIntLargeNsplit}
\end{align}
where $\Gamma_\text{LO}[\Psi,\mathcal{G}]\sim N^1$, $\GammaInt_\NLO[\Psi,\mathcal{G}]\sim N^0$, and so on.  
This constitutes a nonperturbative expansion scheme, which does not rely on small couplings. In this work, we consider~\eqref{eq:GammaIntLargeNsplit} to NLO, and then set $N=1$ to obtain an approximation for the effective action of the single-component Bose gas. At LO one finds
\begin{align} 
    &\GammaInt_\text{LO}[\Psi, \mathcal{G}]
    =
    \int_{x} \bigg \lbrace 
    \frac{g_0}{8} \big[\Psi^\alpha(x) \sigma_1^{\alpha\beta} \Psi^\beta(x) \big]^2
    +\frac{g_N}{4}\sigma_1^{\alpha\beta} \sigma_1^{\gamma\delta}
\notag
\\[0.35em]
    & \hspace{2.5em}\times
       \mathcal{G}^{\alpha\beta}(x,x) 
    \bigg[\Psi^\gamma(x)\Psi^\delta(x)+ \frac{1}{2} \mathcal{G}^{\gamma\delta}(x,x)\bigg]   
    \bigg \rbrace
\, ,
\label{eq:GammaIntLO}
\end{align}
which exclusively contains local terms with a single vertex. Up to a prefactor, $\GammaInt_\text{LO}[\Psi,\mathcal{G}]$ corresponds to the Popov theory in which the anomalous propagator $\tilde G$ is neglected~\cite{andersen2004theory}. UV-divergent vacuum bubbles such as~\eqref{eq:barPvacp} do not appear in~\eqref{eq:GammaIntLO}. Therefore, all counterterms vanish at LO.

At NLO, the following terms contribute:
\begin{align} 
    &\GammaInt_\NLO[\Psi, \mathcal{G}] 
    =\int_{x} \bigg\lbrace
    \frac{\delta g_A}{4}
        \bigg[
            \sigma_0^{\alpha\beta}\sigma_0^{\gamma\delta}
            -
            \sigma_3^{\alpha\beta}\sigma_3^{\gamma\delta}   
        \bigg]
\notag
\\[0.5em]
    & \hspace{0.5em}
    \times
    \mathcal{G}^{\alpha\beta}(x,x)
    \bigg[
            \Psi^\gamma(x)\Psi^\delta(x)+\frac{1}{2} \mathcal{G}^{\gamma\delta}(x,x)
        \bigg]
    \bigg\rbrace
\notag
\\[0.5em]
    & \hspace{0.5em}
    +\frac{1}{2} \tr \left[\ln B \right] 
    +\frac{1}{2} \tr \left[HB ^{-1}\right]
\, ,
\label{eq:GammaIntNLO}
\end{align}
where we employ the notation $\tr\big[A\big]\equiv\int_x A(x,x)$. Here, the trace terms resum an infinite series of loop contributions of the form given in Fig.~\ref{fig:SE_NLO_sketch}. They contain the function
\begin{align}
    B(x,y) 
    &= 
    \delta(x-y) + \Pi (x, y)
\, , 
\label{eq:rsf:B}
\end{align}
together with
\begin{align}
    \Pi(x, y)
    = 
    \frac{g}{2} \sigma_1^{\alpha\beta} \mathcal{G}^{\beta\gamma} (x, y) \sigma_1^{\gamma\delta} \mathcal{G}^{\delta\alpha} (y, x)
\, .
\label{eq:rsf:Pi}
\end{align}
Field-dependencies enter via the function
\begin{align}
\label{eq:rsf:H}
    H (x, y) 
    = 
    g \Psi^\alpha(x) \sigma_1^{\alpha\beta } \mathcal{G} ^{\beta \gamma} (x, y) \sigma_1^{\gamma\delta }\Psi^\delta(y)
\, .
\end{align}
We remark that we have dropped the counterterm $\delta g_N$ from $\GammaInt_\NLO[\Psi,\mathcal{G}]$ since it receives no LO contributions.

The inverse of the function $B(x,y)$ is given by 
\begin{align}
    B^{-1} (x,y) = \delta(x-y) - I (x,y)
\, , 
\label{eq:rsf:Binv}
\end{align}
which involves the resummed bubble chain $I$, which satisfies the recursive relation
\begin{align}
    I (x, y) 
    &= 
    \Pi (x, y) - \int_{z} \Pi (x, z) I(z, y)
\, . 
\label{eq:rsf:I}
\end{align}
Furthermore, in the condensed phase, the following combination of the above functions appears in the self-energy:
\begin{align}
    Q (x,y) &= 
    \int_{z,u} 
    B^{-1}(x, z) H(z,u) B^{-1}(u, y)
\, .
\label{eq:rsf:Q} 
\end{align}
Each of the functions~\eqref{eq:rsf:B}--\eqref{eq:rsf:Q} is time-ordered, in the sense that it is symmetric under the exchange of its arguments, e.g.~$\Pi(x,y)=\Pi(y,x)$. 
By computing the propagator derivative~\eqref{eq:SEcc_def} of $\GammaInt_\text{LO}[\Psi,\mathcal{G}]$ and $\GammaInt_\NLO[\Psi,\mathcal{G}]$, we obtain the corresponding self-energy components~\eqref{eq:SE_and_SEt_fromSEcc} that are required to solve the propagator equations~\eqref{eq:PE_normal} and~\eqref{eq:PE_anom}.
Since the solutions \mbox{$\Psi(x) \equiv \Psi$} and \mbox{$\bar{\mathcal{G}}(x,y)=\bar{\mathcal{G}}(x-y,0)$} are translation invariant, we give the resulting self-energies in momentum-space:
\begin{align}\label{eq:SE_LO}
    \Sigma_\text{LO} (p) 
    =\,& 
    - g_N n   
\, , 
\\[0.3em] 
    \SigmaNLO (p) 
    =\,&  
    - g n 
    + g|\Psi|^2 I (p) 
\notag
\\[0.3em]    
    &+ 
    g \int_{q} G (q) \big[ I (p-q) + Q (p-q)\big]
\, ,
\label{eq:SE_NLO}
\\[0.3em] 
    \tildeSigmaNLO (p) 
    =\,  
    &- g_A \tilde n 
    + 
    g\Psi^2 I (p) 
\notag
\\[0.3em]
    &+  
    g \int_{q} \tilde G (q) \big[ I (p-q) + Q (p-q)\big] 
\, . 
\label{eq:SEt_NLO}
\end{align}

The expression~\eqref{eq:GammaIntLO} for $\GammaInt_\text{LO}[\Psi,\mathcal{G}]$ contains only the normal propagator such that \mbox{$\tilde\Sigma_\text{LO}(p)=0$}. Its NLO contribution~\eqref{eq:SEt_NLO} contains the anomalous average~\cite{andersen2004theory}
\begin{align}
    \tilde n\equiv\Psi^2+ \int_q \tilde G(q)
\, .
\label{eq:tildennc_def}
\end{align}
The required momentum-space expressions of the functions~\eqref{eq:rsf:Pi}--\eqref{eq:rsf:Q} are given by 
\begin{align}
    \Pi(p)
    &=
    g \int_{q} \left[
	   G^*(q) G(p-q) + \tilde G^*(q) \tilde G(p-q) 
    \right]
\, , 
\label{eq:rsf:Pi_p}
\\[0.9em]
    H(p) &= 2 g \, \text{Re} \left[|\Psi|^2  G(p) + \Psi^{2} \tilde G^*(p) \right]
\, ,
\label{eq:rsf:H_p}
\\[0.9em]
    I(p) &= \frac{\Pi(p)}{1+\Pi(p)}
\, , 
\quad
    Q(p) = \frac{ H(p)}{\big[1+ \Pi(p) \big]^2}
\, .
\label{eq:rsf:I_Q_p}
\end{align}
These functions are real and invariant under the reflection of $p$ due to~\eqref{eq:G_and_Gt_symmetry}. 
In the following, we choose $\Psi \in \mathbb{R}$ by virtue of a global $U(1)$-rotation. 
The self-energy expression~\eqref{eq:SEt_NLO} and the anomalous propagator equation~\eqref{eq:PE_anom} then imply that $\tilde G (p)$ and $\tilde \Sigma (p)$ are real as well. 

\subsection{Chemical potential and renormalisation at NLO}\label{ssec:fieldEqAndMu}
Since $\Psi$ does not appear linearly in $\GammaInt[\Psi,\mathcal{G}]$, the field equation~\eqref{eq:FE_hom} always allows the solution $\Psi=0$. Below $T_c$ this solution is forbidden due to number conservation~\eqref{eq:numCons}, which we use to determine $\Psi$ at fixed density $n$. The chemical potential $\mu$ is then fixed by the field equation. At NLO one obtains
\begin{align}
    \mu_\text{LO} &= g_0 \Psi^2 +  g_N \nnc
\, ,
\label{eq:mu_LO}
\\[0.9em]
    \mu_\NLO 
    &= 
    g \nnc + g_A \tnnc - g \int_{q} 
    \left[
	   G(q)+ \tilde G(q)
    \right] I (q)
\, .
\label{eq:mu_NLO}
\end{align}
For completeness, we also give the expression \mbox{$\mu+\Sigma(p)$}, which appears in the propagator equations of motion~\eqref{eq:PE_normal} and~\eqref{eq:PE_anom},
\begin{align}
    \mu_\text{LO}
    +\mu_\NLO
    &+\Sigma_\text{LO} (p)
    +\Sigma_\NLO (p)    
    =
    -(g_N+g)  \Psi^2 
\notag
\\[0.6em] 
    + g_A \tilde n 
    &+ g \Psi^2 I (p) 
    + g \int_{q} \bigg \lbrace G (q) 
    \big[ I (p-q)
    - I (q)   
\notag
\\ 
    & + Q (p-q)\big]  
    - \tilde G(q) I (q) \bigg \rbrace 
\, , 
\label{eq:muPlusSE_atNLO}    
\end{align}
where we used $g_0=g_A$ and $\tilde n = \Psi^2+\tnnc$.

The remaining ingredient for the equations of motion at NLO is the value of the approximation-dependent normal counterterm $\delta g_N$. It is obtained from the renormalisation condition~\eqref{eq:RCforIndComps} for which we require the function $\Lambda^{\alpha\beta,\gamma\delta}(x,y;z,u)$ given by~\eqref{eq:4kernelSimp}. At vanishing external momenta we find
\begin{align} 
    \Lambda&{}_\text{LO}^{\alpha\beta, \gamma \delta} (p=0) 
    = 
    \, g_N \sigma_1 ^{\alpha\beta} \sigma_1 ^{\gamma\delta} 
    \, , 
\label{eq:LambdaLO}
\\[0.5em] 
    \Lambda&{}_\NLO^{\alpha\beta, \gamma \delta} (p=0) 
    = \,
    g\sigma_1 ^{\alpha\beta} \sigma_1^{\gamma\delta} 
    +
    g_A \big(\delta^{\alpha \beta} \delta^{\gamma \delta}- \sigma_3^{\alpha \beta} \sigma_3^{\gamma \delta} 
    \big)
\notag 
\\[0.25em] 
    &-
    g\Big[ \sigma_1 ^{\alpha\gamma} \sigma ^{ \beta \delta}_1  
    +
    \sigma_1 ^{\alpha\delta} \sigma ^{ \beta \gamma}_1 \Big] I(0)
    - 
    g^2 \int_{q} \bigg\lbrace
    \mathcal{G} ^{\alpha\beta}(q) B ^{-1}(q) 
\notag 
\\  
    & \times 
    B ^{-1}(q)
    \Big[
        \mathcal{G}^{\gamma\delta}(q)
        +
        \mathcal{G}^{\delta\gamma}(q)
    \Big] \bigg\rbrace
\, ,
\label{eq:LambdaNLO}
\end{align}
from which \mbox{$\Lambda_{N}^{\text{LO}}(p=0)=g_N$} follows for any state. 

In vacuum, the NLO contribution simplifies due to~\eqref{eq:cyclVacLoopsVanish} which implies \mbox{$\Pi_\vac(p) = I_\vac(p)=0$} and therefore \mbox{$B_\vac(p) = 1$}. 
From~\eqref{eq:LambdaNLO} we recover the exact result~\eqref{eq:LambdaAvac} for the anomalous component along with \mbox{$\Lambda_{N}^\NLO(p=0)\big\vert_\vac=g  - g^2 \bar P_\vac(0)$}. The renormalisation condition~\eqref{eq:RCforIndComps} then yields the normal counterterm
\begin{align}
    \delta g^\NLO_{N} = g ^2 \frac{\UVco}{4 \pi^2}
\, .
\label{eq:deltagNresultNLO}
\end{align}
Our results for the chemical potential, along with the counterterms $\delta g_A$ and $\delta g_N$, ensure that no UV-divergences appear in the equations of motion, as shown in appendix~\ref{app:UV}.

\subsection{HFB approximation}\label{ssec:HFB} 
The HFB approximation corresponds to the first-order weak-coupling expansion~\eqref{eq:GammaHFB_cc} of $\GammaInt[\Psi,\mathcal{G}]$. Equivalently, it is obtained by dropping all loop corrections from the functions~\eqref{eq:SE_NLO},~\eqref{eq:SEt_NLO} and~\eqref{eq:LambdaNLO}.
This results in the momentum-independent self-energies
\begin{align}
    \tilde \Sigma_\HFB = - g_A \tilde n \, ,
\quad
    \mu_\HFB + \Sigma_\HFB = - 2 g \Psi^2 - \tilde  \Sigma_\HFB \, .
\label{eq:SE_HFB}
\end{align}
In this case, the propagator equations~\eqref{eq:PE_normal} and~\eqref{eq:PE_anom} reduce to
\begin{align}
    G (p) = \frac{i p_0 + \xi_{\bm{p}}}{p_0^2 + \epsilon^2_{\bm{p}}}
\, , \quad
    \tilde G (p) = \frac{\tilde \Sigma_\HFB}{p_0^2 + \epsilon^2_{\bm{p}}}
\, ,
\label{eq:HFBpropagators}
\end{align}
where we have defined $\xi_{\bm{p}} \equiv \bm{p}^2 + 2 g \Psi^2 + \tilde \Sigma_\HFB$ and the HFB dispersion relation 
\begin{align}
    \epsilon_{\bm{p}}
    =
    \sqrt{(\bm{p}^2+2g\Psi^2)(\bm{p}^2+2g\Psi^2+2\tilde\Sigma_\HFB)}
\, .
\label{eq:HFBdispersion}
\end{align}
The HFB dispersion relation has a gap which is absent in the Bogoliubov approximation, where \mbox{$\tilde \Sigma (p) = -g \Psi^2$}, such that the Bogoliubov dispersion relation \mbox{$\epsilon_{\bm{p}} ^\text{B}=\sqrt{\bm{p}^2(\bm{p}^2+2g \Psi^2)}$} is recovered. As discussed in Ref.~\cite{zhang2013conserving} the gapped dispersion~\eqref{eq:HFBdispersion} does not violate the Hugenholtz-Pines theorem, since the 2PI resummed two-point function 
\begin{align}
    \Gamma_2^{\alpha\beta}(\Psi;x,y)=\frac{\delta^2 \Gamma[\Psi,\mathcal{G}(\Psi)]}{\delta\Psi^\alpha(x)\delta\Psi^\beta(y)}
\, , 
\label{eq:2PIresummedGama2} 
\end{align}
is gapless. This is due to the 2PI Ward identities, which guarantee that ordinary Ward identities, such as the Goldstone theorem, hold~\cite{reinosa2007QEDward2PI}. 

The HFB equations~\eqref{eq:HFBpropagators} are solved by self-consistently determining $\Psi^2$ and $\tilde\Sigma_\HFB$. For the latter, we  employ~\eqref{eq:tildennc_def} and~\eqref{eq:SE_HFB} to obtain
\begin{align}
    \tilde \Sigma_\HFB 
    = 
    - g_A \left [\Psi^2+ \tilde \Sigma_\HFB \int_p \frac{1}{p_0^2 + \epsilon^2_{\bm{p}}} \right ]
\, .
\label{eq:GapEq}
\end{align}
Since the HFB self-energies are constant in momentum space, the frequency sum in~\eqref{eq:GapEq} can be carried out and we find
\begin{align}
    \tilde \Sigma_\HFB^{-1} 
    = 
    - \frac{1}{g \Psi^2} \left( 
        1+g\int_{\bm{p}} 
        \left \lbrace
            \frac{n_B(\epsilon_{\bm{p}})+\frac{1}{2}}{\epsilon_{\bm{p}}}
            -
            \frac{1}{2\bm{p}^2}
        \right \rbrace 
    \right) 
\, ,
\label{eq:tsigmaHFB}
\end{align}
where $n_B(\epsilon_{\bm{p}})\equiv1/(e^{\beta\epsilon_{\bm{p}}}-1)$. The last term in the integral is due to the counterterm $\delta g_A$ and removes the UV divergence associated with the quantum-half. We remark that the gap equation~\eqref{eq:tsigmaHFB} is self-consistent since the dispersion $\epsilon_{\bm{p}}$ is a function of $\tilde\Sigma_\HFB$.

To close the HFB equations, we need an expression for $\nnc$ to determine $\Psi^2$ via number conservation~\eqref{eq:numCons}. After rewriting the normal propagator in~\eqref{eq:HFBpropagators} as $G(p)=[1+\tilde \Sigma_\HFB\tilde G (p)]/(- i p_0 +  \xi_{\bm{p}})$, summing the frequencies yields
\begin{align} 
    \nnc
    &= 
    (4 \pi \beta)^{-3/2}\,
    \text{Li}_{3/2}\!\left(\exp\left[{-\beta\left(2g \Psi^2+\tilde \Sigma_\HFB\right)}\right]\right)
\notag
\\
    &+ 
    \tilde\Sigma_\HFB^2
    \int_{\bm{p}} 
    \frac{1}{\epsilon_{\bm{p}}}
    \Bigg[
    \frac{1}{2(\epsilon_{\bm{p}}+ \xi_{\bm{p}})}
    +
    \frac{
	   \epsilon_{\bm{p}} n_B(\xi_{\bm{p}})
	   -
	   \xi_{\bm{p}} n_B(\epsilon_{\bm{p}})
    }
    {\epsilon_{\bm{p}}^2 - \xi^2_{\bm{p}}}
    \Bigg]
.
\label{eq:nncHFB}
\end{align}
The integrals in~\eqref{eq:tsigmaHFB} and~\eqref{eq:nncHFB} are convergent, and we evaluate them numerically. 

\section{Conclusion}\label{sec:conc} 
Having derived the renormalisation of the 2PI effective action for Bose gases beyond Gaussian approximations opens up a wide range of applications. In this work, we compute thermodynamic properties from low to high temperatures including the non-Gaussian scaling behaviour at the phase transition. The calculations are done in Euclidean space-time by taking into account the tower of frequencies appearing at finite temperature. As a consequence, we get the universal as well as non-universal properties that are required for a complete thermodynamic description at all temperatures.

The renormalisation of self-consistent approximation schemes turns out to involve additional counterterms at NLO, which are set by the $s$-wave scattering length in our case. As a consequence of renormalisation, the regulator dependence drops out of physical predictions, thus providing the connection between the parameters of the Hamiltonian and observables. While we have performed the analysis for the single component Bose gas, our procedure generalises to more complex non-relativistic quantum theories, such as spinor Bose gases that have so far mostly been limited to Gaussian approximations~\cite{phuc2011effects, Kawaguchi2012finiteTemp}.  

Moreover, self-consistent approximation schemes are important for our understanding of out-of-equilibrium properties of quantum many-body systems~\cite{berges2005bec,Berges2018schwingerBoson, berges2020extracting}. Self-consistency can solve the secularity problem of standard perturbative approaches, which are not uniform in time and become invalid at late times. 
Since renormalisation can still be performed in vacuum also for nonequilibrium applications, our procedure directly applies. This is expected to lead to new insights since alternative non-perturbative approximation schemes in quantum field theory out of equilibrium are scarce.  

\begin{acknowledgments} 
We thank Louis Yussios, Laura Batini and Jendrik Marijan for helpful discussions. This work is part of and
funded by the Deutsche Forschungsgemeinschaft (DFG,
German Research Foundation) through the Collaborative
Research Centre, Project-ID No. 273811115, SFB 1225
ISOQUANT, and Germany’s Excellence Strategy EXC
2181/1–390900948 (the Heidelberg STRUCTURES Ex-
cellence Cluster).
\end{acknowledgments}

\appendix 

\section{Implementation}\label{sec:technicalImplementation}

\subsection{Iterative solver}
Beyond the HFB approximation, it is difficult to find analytical expressions for most loop integrals appearing in the self-energies. 
We compute them numerically using a finite number of spatial momenta and frequencies, which provide ultraviolet and infrared cutoffs $\UVco=|\bm{p}|_\text{max}$ and $\IRco=|\bm{p}|_\text{min}$. 
We solve the propagator equations~\eqref{eq:PE_normal} and~\eqref{eq:PE_anom} iteratively with the ansatz  
\begin{align}
    \Psi^{(i+1)}
    &=
    (1-\alpha)\Psi^{(i)}+\alpha\sqrt{n-\nnc^{(i)}}
\, ,
\\[0.9em]
    G^{(i+1)} (p)
    &=
    (1-\alpha)
    G^{(i)} (p)
\notag
\\[0.5em]
    &\hspace{-2em}+\alpha
    \frac{i p_0  + \bm{p}^2 - \mu  - [\Sigma^{(i)}(p)]^{*}}
    {|-i p_0  + \bm{p}^2 - \mu  - \Sigma^{(i)}(p)|^2-|\tilde \Sigma^{(i)}(p)|^2}
\, ,
\\[0.9em]
    \tilde G^{(i+1)} (p)
    &=
    (1-\alpha)
    \tilde G^{(i)} (p)
\notag
\\[0.5em]
    &\hspace{-2em}+
    \alpha
    \frac{\tilde\Sigma^{(i)}(p)}
    {|-i p_0  + \bm{p}^2 - \mu - \Sigma^{(i)}(p)|^2-|\tilde \Sigma^{(i)}(p)|^2}
\, ,
\label{eq:iteration}
\end{align}
where the first line ensures number conservation according to~\eqref{eq:numCons}. In~\eqref{eq:iteration} we used $\Sigma^{(i)}\equiv\Sigma[G^{(i)}, \tilde G^{(i)}, \Psi^{(i)}]$ and similarly for $\tilde \Sigma^{(i)}$.  Our starting point for the iteration 
\begin{align}
    \!\!\!\Psi^{(0)}=\Psi_\HFB
    \, ,
    \, \ 
    \Sigma^{(0)} (p)=\Sigma_\HFB
    \, ,
    \, \ 
    \tilde\Sigma^{(0)} (p)=\tilde\Sigma_\HFB
    \, ,
\label{eq:iterationStartHFB}
\end{align}
can be obtained by iteratively solving the HFB equations~\eqref{eq:SE_HFB} and~\eqref{eq:nncHFB} along with~\eqref{eq:numCons} for the numbers $\tilde \Sigma_\HFB$ and $\Psi_\HFB$. The mixing parameter $0<\alpha\leq1$ helps stabilising the iteration procedure in non-perturbative parameter regimes, such as close to $T_c$, where we employ $\alpha=0.1$ to achieve convergence.

\subsection{Momentum integrals}
We consider functions $A (p)$ and $B (p)$ in momentum space which satisfy $A^*(p)=A(-p)$. After choosing $\Psi$ real, this is the case for all the two-point functions considered in this work. We make use of isotropy, i.e., $A(p)=A(p_0, |\bm{p}|)$, such that $A^*(p_0,|\bm{p}|)=A(-p_0,|\bm{p}|)$. Convolutions can therefore be computed via positive frequencies only:
\begin{align}
    \int_{q}A(p+q) B&(q) =  \
      T\sum_{q_0=0}^{\Lambda_\omega - p_0} \big[A (p_0+q_0) * B ( q_0) \big] \big(|\bm{p}|\big)
\notag
\\
    +\,  &T\sum_{q_0>0}^{p_0} \big[A(p_0- q_0) * B^* (  q_0)\big] \big(|\bm{p}|\big)
\notag
\\[0.15em]
    +\,  &T\!\sum_{q_0>p_0}^{\Lambda_\omega} 
    \big[A^* (q_0 - p_0) * B^* ( q_0) \big]
    \big(|\bm{p}|\big)
\, . 
\label{eq:unlabelled1}
\end{align}
Here, we have used a frequency cutoff $\Lambda_\omega$ to make the numerical computations tractable.
For spatial convolutions in terms of radial momenta, we employ 
\begin{align}
    \!\!\!\!\big[A(k_0) * B(q_0) \big](|\bm{p}|)
    & \equiv  
    \frac{1}{4 \pi^2} \frac{1}{|\bm{p}|} 
    \int_{0}^{\infty} d|\bm{k}| \, |\bm{k}| \, A(k_0,|\bm{k}|)\,  
\notag
\\[0.3em]
    &\times 
    \int_{||\bm{p}|-|\bm{k}||}^{|\bm{p}|+|\bm{k}|} 
    d|\bm{q}| \, |\bm{q}| \, B(q_0,|\bm{q}|)
\, ,
\end{align}
where the UV and IR cutoffs enter through 
\begin{align}
    A(k_0, |\bm{k}|> \UVco)=A(k_0, |\bm{k}|< \IRco)=0
    \, , 
\end{align}
and similarly for $B(q_0, |\bm{q}|)$. 

Frequency sums typically result in exponential decay as a function of spatial momenta\footnote{Apart from quantum-half like contributions, which are addressed by renormalisation.}. 
Therefore, by inspection of the free propagator \mbox{$G_{0}^{-1} (p)=-i p_0 +  \bm{p}^2-\mu$}, we require the convergence condition \mbox{$\Lambda_\omega \gg \UVco^2$} to properly resolve the UV behaviour. 
Numerically, it is implemented as \mbox{$\Lambda_\omega=K_\omega \UVco^2$} with $K_\omega$ chosen between $10$ and $100$. 
To benchmark the ultraviolet convergence, we have used perturbative loop integrals, such as the ones mentioned in appendix~\ref{app:UV}. 
In contrast to standard perturbative computations, loop integrals involving self-consistently resummed propagators are generally finite in the infrared~\cite{berges2005renormalisedThermo}. 

In practice, we use logarithmically spaced values of $|\bm{p}|$ making sure that all physical scales such as the healing length $\xi_\text{h}=(gn)^{1/2}$ and the thermal wavelength $\lambdaTh$ are separated from the cutoffs by at least one order of magnitude. For computations at criticality we lower to infrared cutoff until the scaling regime of the zero-frequency mode $G(p_0=0, \bm{p})$ is entered. 

\subsection{Non-condensed phase}\label{sec:symmPhase}
For computations at $T\geq T_c$, the equations of motion simplify due to $\Psi=\tilde G(p)=\tilde \Sigma(p)=0$. 
The remaining equation of motion for the normal propagator~\eqref{eq:PE_normal} becomes
\begin{align}
    G ^{-1}(p) = -i p_0 + \bm{p}^2 - \bar\mu - \big[\Sigma(p)-\Sigma(0) \big] 
\, .
\label{eq:symmPEshifted}
\end{align}
In the non-condensed phase, the field equation~\eqref{eq:FE} cannot be used to determine the chemical potential $\mu$. Instead, it is solved by the trivial solution $\Psi=0$ and we use the shifted chemical potential
\begin{align}
    \bar \mu &= \mu + \Sigma (0)
    \, ,
\label{eq:shiftedChemPot}
\end{align}
as a Lagrange multiplier for the number density $n$. 
If an approximation leads to a constant self-energy $\Sigma(p)$, the form of the propagator~\eqref{eq:symmPEshifted} becomes non-interacting with $\mu$ replaced by $\bar\mu$. 
Therefore, the HFB and LO approximations predict the same critical behaviour as the ideal Bose gas. 

Setting $\bar{\mu}=0$ allows us to work directly at the critical point, where $G(p=0)=0$ such that the correlation length diverges~\cite{baym2001transitionTemperatureBose}. 
Since the divergence occurs only at $p_0=0$, universal quantities such as the anomalous dimension $\eta$ and the slope $c$ of the shift in $T_c$ are entirely determined by the zero-frequency mode. 
They can thus be computed in a dimensionally reduced theory, where frequencies with $p_0>0$ are neglected.

\subsection{Particle number}
The density of non-condensed atoms is given by~\eqref{eq:nnc_fromG}, which reads in momentum space 
\begin{align}
    \nnc = 
    \lim_{\delta x_0 \searrow 0 } \int_{p} e^{i p_0 \delta x_0}G (p)
\, .
\label{eq:nnc_from_Gp}
\end{align}
Bose-Einstein condensation occurs when this integral fails to account for the entire atomic density $n$. For computations at NLO, it is convenient to write the integral~\eqref{eq:nnc_from_Gp} as
\begin{align}
    \nnc=\nncFree+\nncInt 
\, ,
\label{eq:nnc_split}
\end{align}
where we have isolated the non-interacting part 
\begin{align}
    \nncFree&=
    \lim_{\delta x_0 \searrow 0 } \int_{p} e^{i p_0 \delta x_0}G_0 (p)
    =
    \lambdaTh^{-3}\text{Li}_{3/2}(e^{\beta \mu})
\, ,
\label{eq:nnc_0_fullComputation}
\end{align}
which vanishes at zero temperature. For the interacting part, the limit can be taken inside the integral. By rearranging the propagator equation~\eqref{eq:PE}, we obtain the integral
\begin{align}
    \nncInt &= \int_{p} G_0 (p) \left[ 
        \Sigma (p) G (p)
        + 
        \tilde\Sigma^* (p) \tilde G (p) 
    \right]
\, ,
\label{eq:delta_n}
\end{align}
which we evaluate numerically when working at NLO. 

\section{Vacuum self-energy and vertices}\label{app:vacuum}
In the following, we evaluate the exact expressions~\eqref{eq:Gvac},~\eqref{eq:LambdaAvac},~\eqref{eq:LambdaN_vac_exact_diagrams} and~\eqref{eq:GammaInt1122_vac} for $G_\vac$, $\Lambda_\vac$ and $\delta^4\GammaInt/\delta\Psi^4\big\vert_\vac$, which we employed in section~\ref{sec:renorm} to renormalise the two-body coupling in vacuum. We start by deriving the fact that cyclic vacuum loops vanish as depicted in~\eqref{eq:cyclVacLoopsVanish}.

In the non-condensed phase, an arbitrary loop integral can be written as
\begin{align}\notag
    P_{m,\bar m}(G;p_1,  &\dots , p_{m+ \bar m}) 
    =
    \int_{q} 
    \Big[
    G(p_1+q)
    \dots
    G(p_m+q)
\\ 
    &\times G ^*(p_{m+1}+q)
    \dots
    G ^*(p_{m+ \bar m}+q)
    \Big]
\, .
\label{eq:PsymmComponents} 
\end{align}
The frequency sum contained in ${\int_{q} \equiv \int \frac{\mathrm{d}^3\bm{q}}{(2 \pi)^3} T \sum_{q_0}}$ generally splits into a zero and non-zero temperature part. The zero temperature part can be obtained by replacing the frequency sum by its zero temperature limit:
\begin{align}
	T \sum_{q_0} \longrightarrow \int_{- \infty}^{\infty} \frac{\mathrm{d} q_0}{2\pi}
\, .
\label{eq:T0freqInt}
\end{align}
To investigate the impact of~\eqref{eq:T0freqInt}, we consider the loop integral~\eqref{eq:PsymmComponents} in terms of the non-interacting propagator $G_0^{-1}(p)=-i p_0 + \bm{p}^2 - \mu$ with $\mu\leq 0$:
\begin{align} 
    &P_{m,\bar m}(G_0;p_1,  \dots , p_{m+ \bar m})\big\vert_{T=0}
    =
    \int \frac{\mathrm{d}^3 \bm{q}}{(2\pi)^3} 
    \int_{- \infty}^{\infty} \frac{\mathrm{d} q_0}{2\pi} \Bigg\lbrace
\notag
\\ 
    &\hspace{2em}\times\bigg[  
        \prod_{j=1}^m\frac{1}{-i q_0 - i p_{0,j} + (\bm{p}_j+\bm{q})^2 -\mu}
    \bigg] 
\notag 
\\
    &\hspace{2em} \times\bigg[ \,
        \prod_{k=m+1}^{m+\bar m}\frac{1}{i q_0 + i p_{0,k} + (\bm{p}_k+\bm{q})^2 -\mu}
    \bigg]\Bigg\rbrace.
\label{eq:PnmG0} 
\end{align}
By means of analytic continuation, we can extend the $q_0$-integration contour to the complex plane. For $m+\bar m\geq 2$ we may close it in the upper half-plane, since the integrand decays faster than $|q_0|^{-1}$ for $|q_0|\rightarrow \infty$. As a consequence, one finds that only the poles at $q_0 = -p_{0,k}+i[(\bm{p}_k + \bm{q})^2 -\mu]$ contribute to the integral. For $\bar m=0$, there are no such poles, and the zero-temperature part of cyclic integrals vanishes:
\begin{align}
    P_{m, 0}(G_0; p_1, \dots, p_{m}) \big\vert_{T=0} = 0
\, ,
\label{eq:CyclicLoopsVanishAtT0}
\end{align}
as depicted in~\eqref{eq:cyclVacLoopsVanish} for the vacuum case. Non-cyclic loops, which have $\bar m >0$, are in general non-vanishing, since they always lead to poles in each half-plane. 

A crucial consequence of~\eqref{eq:CyclicLoopsVanishAtT0} is the fact that the vacuum self-energy vanishes:
\begin{align}
    \Sigma(G_\vac)\equiv\Sigma_\vac=0
\, .
\end{align} 
Therefore, the vacuum propagator 
\begin{align}
    G_\vac(p)=(-ip_0+\bm{p}^2)^{-1}
\, , 
\label{eq:app:Gvac}
\end{align}
solves the equation of motion~\eqref{eq:symmPEshifted} with $\bar \mu =0$ such that the non-interacting dispersion relation is recovered.
To show this, we write the normal self-energy as
\begin{align}
    \raisebox{-0.8em}{\hbox{\includegraphics[scale=\downScale]{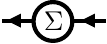}}}
    \ = \
    \raisebox{-1em}{\hbox{\includegraphics[scale=\downScale]{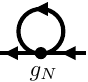}}}
    \ + \,
    \vcenter{\hbox{\includegraphics[scale=\downScale]{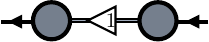}}}
\, ,
\label{eq:VacSE_1}
\end{align}
where the first term equals $g_N n_\vac=0$ in vacuum\footnote{Here, we have discarded a momentum independent divergent term from the one-loop diagram. Equivalently, this term drops out of~\eqref{eq:app:Gvac} by the subtraction in~\eqref{eq:symmPEshifted} together with the renormalisation condition $\bar\mu_{\vac}=0$.}. 
In vacuum, these terms vanish as well: Since all contributions to the self-energy~\eqref{eq:VacSE_1} must be one-particle irreducible, we can write in the absence of a condensate~\cite{luttingerWard1960,cornwallJackiwTomboulis1974effectiveActionForComposite},
\begin{align}
	\vcenter{\hbox{\includegraphics[scale=\downScale]{Diagrams/SigmaVac_blobs_1.pdf}}}
    \ = \
    \raisebox{-0.725em}{\hbox{\includegraphics[scale=\downScale]{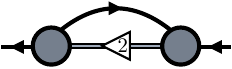}}}
\, , 
\label{eq:VacSE_2}
\end{align}
whose simplest contribution is obtained by replacing each blob by a single vertex. The resulting diagram vanishes in vacuum due to cyclicity~\eqref{eq:cyclVacLoopsVanish}:
\begin{align}
	\vcenter{\hbox{\includegraphics[scale=\downScale]{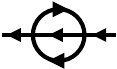}}} 
    \, \big\vert_\vac
    \ = \ 0
\, .
\label{eq:vacSEsunset}
\end{align}
Cyclicity persists at higher orders: Both attempts to avoid cyclicity in~\eqref{eq:VacSE_2}, either by separating propagators,
\begin{align}
	\raisebox{-0.725em}{\hbox{\includegraphics[scale=\downScale]{Diagrams/SigmaVac_blobs_2.pdf}}}
    \ \rightarrow \
    \raisebox{-0.725em}{\hbox{\includegraphics[scale=\downScale]{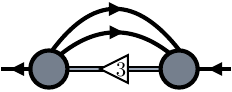}}}
\, , 
\label{eq:vacSE_3}
\end{align}
or by further decomposition of subdiagrams, 
\begin{align}
    \raisebox{-0.725em}{\hbox{\includegraphics[scale=\downScale]{Diagrams/SigmaVac_blobs_2.pdf}}}
    \ \rightarrow \
    \vcenter{\hbox{\includegraphics[scale=\downScale]{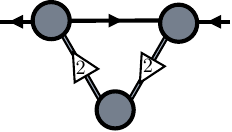}}}
    \label{eq:vacSE_4}
\end{align}
fail. 
It is straightforward to see that further iterations or combinations of the manipulations~\eqref{eq:vacSE_3} and~\eqref{eq:vacSE_4} do not avoid cyclicity either. 
Therefore, the vacuum self-energy vanishes identically.
Moving on to $\Lambda^\vac_A$, we write
\begin{align}
    \vcenter{\hbox{\includegraphics[scale=\downScale]{Diagrams/LambdaA.pdf}}}
    \ = 
    \raisebox{-1.5em}{\hbox{\includegraphics[scale=\downScale]{Diagrams/vertex_gA.pdf}}}
     +  \,
    \vcenter{\hbox{\includegraphics[scale=\downScale]{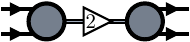}}}
    \, +  
    \vcenter{\hbox{\includegraphics[scale=\downScale]{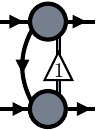}}}
\label{eq:LambdaA_splitIntoBlobs}
\end{align}
where the first term is obtained by cutting both anomalous propagator lines from the last diagram in~\eqref{eq:GammaHFB_diag}. 
Regarding the loop terms in~\eqref{eq:LambdaA_splitIntoBlobs}, we included both ways of grouping their vertices into two subdiagrams. 
The second term vanishes in vacuum through~\eqref{eq:cyclVacLoopsVanish}. 
Its cyclic structure persists under manipulations such as~\eqref{eq:vacSE_3} and~\eqref{eq:vacSE_4}. 
The first one vanishes as well, which can be seen by rewriting it as
\begin{align}
    \vcenter{\hbox{\includegraphics[scale=\downScale]{Diagrams/LambdaA_blobs_1.pdf}}}
    \ =& \ \,
    \vcenter{\hbox{\includegraphics[scale=\downScale]{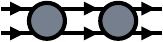}}}
\notag
\\[0.3em]
    &+ \
    \raisebox{-0.96em}{\hbox{\includegraphics[scale=\downScale]{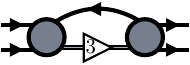}}}
\: .
\label{eq:LambdaA_2}
\end{align}
The first diagram on the right-hand side can be disconnected into a left and right piece by cutting the two centre propagators. 
It is therefore absent from $\Lambda_A$, which only receives contributions from two-particle irreducible diagrams in $\GammaInt[\Psi,\mathcal{G}]$ via~\eqref{eq:LambdaDef}. 
The second diagram in~\eqref{eq:LambdaA_2} vanishes in vacuum due to cyclicity. 
Therefore, only the single-vertex term in~\eqref{eq:LambdaA_splitIntoBlobs} survives in vacuum and~\eqref{eq:LambdaAvac} follows.
Repeating the above steps for the component $\Lambda_N$, we write
\begin{align} 
    \vcenter{\hbox{\includegraphics[scale=1]{Diagrams/LambdaN.pdf}}}
    \ = \,
    \raisebox{-1.5em}{\hbox{\includegraphics[scale=\downScale]{Diagrams/vertex_gN.pdf}}}
    \, + 
    \vcenter{\hbox{\includegraphics[scale=\downScale]{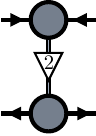}}}
    + \,
    \raisebox{-1em}{\hbox{\includegraphics[scale=\downScale]{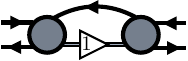}}}
\, ,
\label{eq:LambdaN_1}
\end{align}
where the third term exhibits a cyclic structure and vanishes in vacuum. 
The second one is not forbidden by two-particle irreducibility, and its vacuum contributions are  
\begin{align}
    \vcenter{\hbox{\includegraphics[scale=\downScale]{Diagrams/LambdaN_blobs_1.pdf}}} \, 
    \, \big\vert_\vac =
    -\frac{1}{2}\vcenter{\hbox{\includegraphics[scale=\downScale]{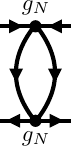}}} \, 
    +\frac{1}{4}\vcenter{\hbox{\includegraphics[scale=0.975]{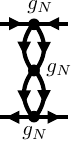}}} \, 
    + \dots
\label{eq:LambdaN_3}
\end{align}
These contributions are all non-cyclic and thus non-vanishing in vacuum, leading to the result~\eqref{eq:LambdaN_vac_exact_diagrams}. 
Finally, we derive~\eqref{eq:GammaInt1122_vac} by considering the terms in $\GammaInt[\Psi,\mathcal{G}]$ that contain four field insertions. 
Along the same lines as for the decompositions~\eqref{eq:LambdaA_splitIntoBlobs} and~\eqref{eq:LambdaN_1}, we write them as
\begin{align}
	\vcenter{\hbox{\includegraphics[scale=0.975]{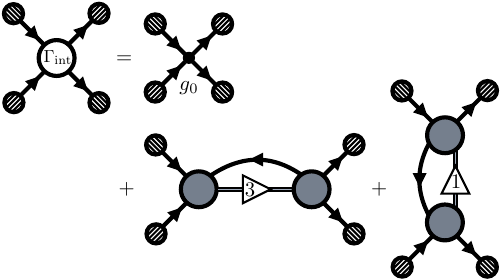}}}
\label{eq:GammaInt1122_1}	
\end{align}
Only the single-vertex term is both two-particle irreducible and non-cyclic. It is therefore the only term contributing in vacuum.

\section{Absence of UV-divergences in the renormalised theory} \label{app:UV}
In this section, we illustrate that the renormalised equations of motion~\eqref{eq:PE_normal} and~\eqref{eq:PE_anom} are finite in the limit \mbox{$\UVco\rightarrow\infty$}. We consider at NLO the UV behaviour of $\Sigma+\mu$ and $\tilde \Sigma$, which are given by~\eqref{eq:SEt_NLO} and~\eqref{eq:muPlusSE_atNLO}. 
For example, using the asymptotic forms of the functions 
\begin{align}
    G^{-1}(p)&\UVasymptotics -ip_0 +\bm{p}^2
\, ,
\\[0.9em]
    \tilde G(p)
    &\UVasymptotics
    \frac{\text{const.}}{p_0^2+(\bm{p}^2)^2}
\, ,
\\[0.9em]
    I(p) 
    &\UVasymptotics 
    g\int_q G^*(q)G(p-q)
\, ,
\\[0.9em]
    Q(p)  
    &\UVasymptotics 
    g\Psi^2  [G(p)+G^*(p)]
\, , 
\label{eq:UV:asymptotics}
\end{align}
one can show that the following integrals are linearly UV-divergent, in the sense that they contain $p$-independent terms proportional to $\UVco$, 
\begin{align}
    \tnnc &= \int_q \tilde G(q) 
\, , 
\label{eq:UV:intnnc}
\\[0.9em]
    \mathcal{I}_I (p) &= \int_q G(p-q)I(q)
\, ,
\label{eq:UV:intI}
\\[0.9em]
    \mathcal{I}_Q (p) &=  \int_q G(p-q)Q(q)
\, .
\label{eq:UV:intQ}
\end{align}
The divergence in~\eqref{eq:UV:intnnc} drops out since only the product $g_A\tnnc$ appears in~\eqref{eq:SEt_NLO} and~\eqref{eq:muPlusSE_atNLO}. 
It is finite in the limit $\UVco \rightarrow \infty$ since $g_A\sim\UVco^{-1}$ due to~\eqref{eq:gAresult}. 
This can be seen explicitly within the HFB-approximation where $g_A\tnnc$ is given by~\eqref{eq:SE_HFB} and \eqref{eq:tsigmaHFB}.

The divergent part of $\mathcal{I}_I(p)$ is momentum independent and is therefore cancelled by the chemical potential, whose solution~\eqref{eq:mu_NLO} in the condensed phase or shift~\eqref{eq:shiftedChemPot} in the non-condensed phase imply that only the quantity $\mathcal{I}_I(p)-\mathcal{I}_I(0)$ is physically relevant~\cite{baym2001transitionTemperatureBose}. 
Moving on to $\mathcal{I}_Q(p)$, according to~\eqref{eq:muPlusSE_atNLO} we need to consider the combination $g \mathcal{I}_Q(p) - \Psi^2\delta g_N$, where the normal counterterm $\delta g_N$ is given by~\eqref{eq:deltagNresultNLO}. 
Plugging in~\eqref{eq:UV:asymptotics} and dropping the finite integral $\int_q G(q)G^*(p-q)$, we obtain
\newpage
\noindent
\begin{align} 
    g \mathcal{I}_Q(p) - \Psi^2\delta g_N 
     \UVasymptotics
    \Psi^2 g^2\left(
         \int_q \Big[ G(q)G(p-q) 
        \Big]
        -
        \frac{\UVco}{4\pi^2}
    \right)
\end{align}
which is UV-finite because
\begin{align} 
    \int_q  G(q)G(p-q) 
    =
    \frac{\UVco}{4\pi^2} + \mathcal{O}\big((\UVco)^0\big)
    \, . 
\end{align}
\bibliography{references}
\end{document}